\documentclass[11pt,preprint,flushrt]{aastex}

\def\eqq#1{Equation~(\ref{#1})}
\newcommand\etal{{\it et al.\/}}
\newcommand\eg{{\it e.g.\ }}
\newcommand\ie{{\it i.e.\ }}

\begin{document}

\slugcomment{CVS \$Revision: 1.19 $ $ \$Date: 2008/08/25 15:39:40 $ $}

\title{Comprehensive Two-Point Analyses of Weak Gravitational Lensing Surveys} 
\author{Gary M. Bernstein}
\affil{Dept. of Physics and Astronomy, University of Pennsylvania,
Philadelphia, PA 19104}
\email{garyb@physics.upenn.edu}

\begin{abstract}
We present a framework for analyzing weak gravitational lensing survey
data, including lensing and source-density observables, plus spectroscopic
redshift calibration data.  All two-point observables are predicted in
terms of parameters of a perturbed Robertson-Walker metric, making
the framework independent of the models for gravity, dark energy, or
galaxy properties.  For Gaussian fluctuations the 2-point model determines the
survey likelihood function and allows Fisher-matrix forecasting.  The
framework 
includes nuisance terms for the major systematic errors: shear
measurement errors, magnification bias and redshift calibration
errors, intrinsic galaxy alignments, and inaccurate theoretical
predictions.   We propose flexible parameterizations of the many
nuisance parameters related to galaxy bias and intrinsic alignment.
For the first time we can integrate many different observables and
systematic errors into a single analysis.
As a first application of this framework, we demonstrate that:
uncertainties in power-spectrum theory cause very minor degradation to
cosmological information content; nearly all useful information
(excepting baryon oscillations) is
extracted with $\approx 3$ bins per decade of angular scale; and the rate
at which galaxy bias varies with redshift substantially influences the
strength of cosmological inference.  The framework will permit careful
study of the interplay between numerous observables, systematic
errors, and spectroscopic calibration data for large weak-lensing surveys.

\end{abstract}


\section{Introduction}
Weak gravitational lensing of background sources can produce
exceptionally strong constraints on cosmological parameters and tests
of General Relativity.  Initial analyses considered the 2-point
correlation function (or, equivalently, power spectrum) of the shear
pattern induced on a single population of background galaxies
\citep{Jordi, Kaiser92, Blandford91}.  A wealth of new statistics,
however, have been suggested as more powerful means to extract
information from weak lensing (WL): cross-power spectra of multiple
source populations with distinct redshift distributions
(a.k.a. ``tomography'') \citep{Hu}; the correlation of shear with
foreground galaxy clusters \citep{JainTaylor}, or more generally the
cross-correlation of lensing shear with the galaxy distribution
\citep{BJ04, Zhang}; joint analyses of density-density,
density-shear, and shear-shear correlations in an imaging survey
\citep{HJ04}; cross-correlation of magnification as well as shear
\citep{Jainmag}; use of the CMB \citep{HO02,HS03} or recombination-era 21~cm
signals \citep{MW07,Pen04,ZZ06} as source planes; cross-correlation of source
density or shear with a distinct spectroscopic galaxy survey
population \citep{Newman, SKZC}; and the use of 3-point statistics
\citep{TJ04} or statistics such as peak counts \citep{Hennawi,Wang,Laura} to
move beyond 2-point information.  Each of these potential innovations
has been individually analyzed and shown to improve cosmological
constraints.  The first goal of this paper is to consider the {\em
  simultaneous} use of all of these observable statistics: can we
forecast the cosmological information that they will yield
collectively in future surveys?  Can we start to develop a framework
in which all these signals could be analyzed simultaneously in a real
experiment? 

In parallel with the increasing variety of proposed WL signals, the
community has identified a series of potential astrophysical and
instrumental non-idealities in WL data which, if ignored, would lead
to substantial systematic errors in the inferred cosmology.  These
include: finite accuracy in our ability to predict the deflecting mass
power spectrum due to nonlinearities \citep{JS97} and baryonic physics
\citep{Zhan06, Jing}; intrinsic alignments (IA) between galaxy shapes
\citep{CroftMetzler} and between galaxy shapes and the local mass distribution
\citep{Hirata2} that are not induced by lensing; multiplicative
``shear calibration'' errors in the derivation of lensing shear from
galaxy images \citep{Ishak04,HTBJ}; additive ``spurious shear'' due
to uncorrected PSF 
ellipticity or other imaging systematics \citep{HTBJ,AmaraRefregier}; 
and errors in
the assignment of redshifts to the source populations
\citep{MaHutererHu}.  The impact of these systematic-error sources on
cosmological inferences have been analyzed by different means, but a
second goal of this paper is to produce a comprehensive forecast that
considers the presence of them all simultaneously. 

Previous work has shown that these multiple sources of information and
systematic in WL surveys can interact in interesting ways.  For
example, in the presence of tomographic data, many systematics are
readily distinguishable from cosmological signals and can hence be
diagnosed and corrected internally to a survey; this approach is
called {\em self-calibration} \citep{HTBJ}.  It has also been shown
that combining galaxy density and lensing correlations can lead to
self-calibration of shear calibration errors \citep{BJ04} and the
uncertainties in galaxy biasing \citep{HJ04, Zhan06}.  Intrinsic
alignments of galaxies can be diagnosed and corrected if tomographic
information is available \citep{KingSchneider}, however this places
substantially greater demands on the precision and accuracy of
redshift assignment than would otherwise be needed \citep{BridleKing}.
These investigations raise important practical questions: will the
self-calibration techniques continue to succeed when we attempt to
simultaneously self-calibrate several different systematic errors?  Do
cross-correlation techniques reduce uncertainties in redshift
distributions to negligible levels, or is it necessary to make a
complete spectroscopic redshift survey of some size to measure
redshift distributions directly \citep{MaBernstein}?  This paper will
present a formalism through which all these questions can be answered,
but we defer to later papers the application of the framework to these
issues. 

A third goal of this work is to describe the constraints by WL in a
language that is not tied to a specific cosmological model.  Most
forecasts for WL survey constraints are done within the context of a
Universe that has homogeneous dark energy with equation of state
$w=w_0 + w_a(1-a)$.  Projecting the WL experiment onto this model
gives concrete predictions, but obscures what the WL is really
measuring.  So the analysis framework presented here will be {\em
  dark-energy agnostic}, meaning that no specific model is assumed.
We will be very explicit about the assumptions made in the analysis
and try to keep them to a minimum.  In fact a great strength of WL
experiments are their ability to test General Relativity itself, so we
seek an analysis method that is general enough to incorporate such
tests. Similar to the approach of \citet{KST}, our analysis results in
constraints on the distance and growth functions $D(z)$ and
$g_\phi(z)$, without reference to the particular dark-energy or
gravity modifications that might cause deviations from $\Lambda$CDM.

In the following section we describe a ``kitchen-sink'' formalism for
WL survey observables that allows the incorporation of all suggested
2-point statistics and very general treatments of nearly all proposed
systematic errors.  In \S\ref{speclike} we give a likelihood function
and Fisher matrix for an unbiased spectroscopic redshift survey of
source galaxies.  Then we briefly describe a software implementation
of the lensing and spectroscopy likelihood calculations.
We describe our model for the evolution of the lensing-potential power
spectrum in \S\ref{powermodel}, and \S\ref{nuisance} we describe
generic models used for the nuisance functions required in the
lensing-survey analysis. In \S\ref{tuneup} we use the implementation
of these methods to investigate the proper choices for the bin sizes
and grid spacings needed to turn the lensing analysis into a tractable
finite-dimensional problem.  Further application of the framework to
survey forecasting will be done in future papers.

An earlier version of this WL analysis formalism was used to generate
forecasts for the Dark Energy Task Force \citep{DETF}, and is
described in an appendix to that report.  

\section{The Weak Lensing 2-Point Likelihood}
\subsection{Observables}
\label{observsec}
We make the assumption that {\em the
  Universe has only weak scalar
  perturbations to a homogeneous and isotropic 4-dimensional metric.}
In this case 
the metric can be 
written in the Newtonian gauge as a perturbed Robertson-Walker metric:
\begin{equation}
ds^2 = (1+2\Psi)dt^2 - a^2(t)(1+2\Phi)\left[d\chi^2 +
  \chi_0^2S_k^2(\chi/\chi_0)(d\theta^2+\sin^2\theta d\varphi^2)\right]
\end{equation}

We assign all mass and sources in the Universe to 
a series of narrow spherical shells centered at redshifts
$a_i=(1+z_i)^{-1}$ for $i\in \{1,\ldots,N_z\}$.    
There is a comoving angular
diameter distance $D_i$ to each shell, and the comoving radial extent of
each shell is $\Delta\chi_i$.  Note that the Robertson-Walker metric
formula for angular-diameter distance is $D=\chi_0 S_k(\chi/\chi_0)$,
where $\chi_0$ is the comoving radius of curvature of the Universe.   
For small values of the curvature $\omega_k\equiv -k/\chi^2_0$ 
we have 
\begin{equation}
\Delta\chi_i \approx \Delta D (1-\omega_k D_i^2/2)
= { D_{i+1} - D_{i-1} \over 2} (1-\omega_k D_i^2/2).
\end{equation}
The Robertson-Walker metric also requires $\Delta\chi_i = \Delta z_i /
h(z_i) = \Delta a_i / a_i^2 h(a_i)$.  In this paper the Hubble
parameter will be written as $H(z)=h(z)H_{100}$, $H_{100}=100\,{\rm
  km}\,{\rm s}^{-1}\,{\rm Mpc}^{-1}$, and all distances will be in
units of $c/H_{100}=2998$~Mpc. 

We assume that the photon sources in a survey will be
divided into a series of {\em sets} $\alpha\in\{1,2,\ldots, N_s\}$.
Note the use of latin
indices for redshift shells, greek for source sets.
We follow \citet{HJ04} by assigning
each source set up to two observables: first its sky-plane density
fluctuations $g_\alpha(\theta,\varphi)$, and second a lensing convergence
$\kappa_\alpha(\theta,\varphi)$.  The convergence $\kappa$ might be
inferred from the shear or flexion \citep{BaconFlex} of galaxies, by a
quadratic estimator on the 
CMB or 21-cm radiation fluctuations, or by any other observable except the
source density.  The sources can be assigned to sets by photometric or
spectroscopic redshift, or even cruder color criteria
\citep{ConnollyJain}, but there could be other criteria such as galaxy
type, or perhaps observation by different instruments.  We demand 
only that the criteria for division of the sources be spatially
homogeneous, and that the division be invariant under application of
gravitational lensing distortion.  For notational convenience we
assign each set a nominal redshift $z_\alpha$, but a set can span a
broad redshift range.  If the sources are discrete objects such as
galaxies, then 
the mean density on the sky of members of each set are denoted
$n_\alpha$.

A source in set $\alpha$ has a probability $p_{\alpha i}$ of lying on
redshift shell $i$.   The collection of galaxies in set $\alpha$ on
shell $i$ will be called the {\em subset} $\alpha i$.  The survey is
assumed to tell us only which {\em set} any individual galaxy belongs to,
but not which {\em subset.}  The $p_{\alpha i}$ are parameters which
must be constrained by the lensing survey data or by additional  
observations, {\it e.g.} a spectroscopic redshift survey.  

When the lensing sources are drawn from a spectroscopic survey (or
when the source is the CMB), then the redshift
probability is known {\it a priori}, and in particular the sets are
probably divided by redshift so that $p_{\alpha i}$ is essentially the
identity matrix.  The formalism can obviously accommodate the
simultaneous analysis of WL samples with varying modes of redshift
assignment. 

Both the source density fluctuation $g_\alpha$ and convergence $\kappa_\alpha$
have a component due to
intrinsic fluctuations plus a component due to gravitational lensing.
Both are also measured as weighted sums over their respective subsets.
We have 
\begin{eqnarray}
\label{density1}
1+g_\alpha(\theta,\varphi) & = & \sum_i p_{\alpha i}\left[1+g^{\rm
    int}_{\alpha i}(\theta,\varphi)\right]
 \left[1 + q_{\alpha i} \kappa^{\rm lens}_i(\theta,\varphi)\right] \\
\kappa_\alpha(\theta,\varphi) & = & \sum_i p_{\alpha i}\left[\kappa^{\rm int}_{\alpha i}(\theta,\varphi) + 
(1+f_{\alpha i}) \kappa^{\rm lens}_i(\theta,\varphi)\right].
\end{eqnarray}
Here we have assigned each subset a {\em magnification bias factor}
$q_{\alpha i}$ and a {\em shear calibration factor} $f_{\alpha i}$.
In a simple flux-limited selection, the magnification bias factor will be
determined by the logarithmic slope of the counts vs flux, and is typically of
order unity.  The shear calibration factor allows for the possibility
that the inferred lensing convergence is mis-measured by some factor
$1+f_{\alpha i}$ due to multiplicative errors in the lensing
methodology, \eg as investigated by \citet{STEP1}.

In the limit $g^{\rm int}\ll 1$ and $\kappa^{\rm lens}\ll 1$,
we can drop the second-order term in Equation~\ref{density1} and write
\begin{eqnarray}
\label{density2}
g_\alpha & = & \sum_i p_{\alpha i}\left[g^{\rm int}_{\alpha i} +
q_{\alpha i} \kappa^{\rm lens}_i\right] \\
\label{kappa2}
\kappa_\alpha & = & \sum_i p_{\alpha i}\left[\kappa^{\rm int}_{\alpha i} + 
(1+f_{\alpha i}) \kappa^{\rm lens}_i\right].
\end{eqnarray}
In this case the equations are linear in all the angular functions $g$
and $\kappa$, so we can decompose them into spherical harmonic
coefficient $g_{\alpha \ell m}$, $\kappa^{\rm lens}_{i\ell m}$,
etc, and Equations~(\ref{density2}) and (\ref{kappa2}) hold
independently for every harmonic $\ell m$.  We will henceforth assume
that the spherical-harmonic decomposition has been executed for all
the angular functions $g$, $\kappa$, and suppress the $\ell m$
indices for brevity.

We note that while $\kappa^{\rm lens}\ll 1$ is a good approximation
over most of the sky, $g^{\rm int}\ll1$ is a poor approximation for
thin density slices on smaller angular scales.  We will forge ahead
nonetheless with the assumption that {\em lensing magnification simply
  adds to the intrinsic density fluctuations}, recognizing that a real
analysis of data 
with magnification bias may require inclusion of the nonlinear
coupling between spherical harmonics that is induced by magnification
bias on highly structured density fields. 

The lensing convergence is determined entirely by the metric if we
make the assumption that light rays are following its null geodesics.
The paths of null geodesics are determined by the {\em lensing
  potential}
\begin{equation}
\phi \equiv {1 \over 2}(\Psi-\Phi).
\end{equation}
For each of our redshift shells we define a projected lensing
potential via
\begin{equation}
\psi_i \equiv 2 \nabla^2_\theta \int_{\Delta\chi} \phi a\, d\chi,
\label{lenspsi}
\end{equation}
where the derivatives are taken with respect to angles on the sky.  We
will generally assume that $\psi$, like the observables, has been
decomposed into spherical harmonics, and we will take the flat-sky
approximation.

With the definition (\ref{lenspsi}), and the adoption of the
weak-lensing limit and Born approximation, the lensing convergence is
\begin{eqnarray}
\label{lenslaw}
\kappa^{\rm lens}_i 
& = & \sum_j A_{ij} {\psi_j
  \over 2a_jD_j}, \\
\label{lenslaw2}
A_{ij} & \equiv & 
\left\{
\begin{array}{cl}
{D_{ij} \over D_i} \approx (1-D_j / D_i) (1-\omega_k
D_i D_j / 2) & z_i>z_j, \\
0 & z_i\le z_j.
\end{array}
\right.
\end{eqnarray}
$D_{ij}$ is the comoving angular diameter distance to $z_i$ as viewed
from $z_j$.
In summary, the observables from the survey are, for each spherical harmonic:
\begin{eqnarray}
\label{observables}
g_\alpha & = & \sum_ip_{\alpha i}
\left[ q_{\alpha i} \sum_j A_{ij} {\psi_j \over 2a_j D_j}
 + g^{\rm int}_{\alpha i } \right] \\
\nonumber
\kappa_\alpha & = & \sum_ip_{\alpha i}
\left[ (1+f_{\alpha i}) \sum_j A_{ij} {\psi_j \over 2 a_j D_j}
 + \kappa^{\rm int}_{\alpha i } \right].
\end{eqnarray}
We reiterate that these equations depend only upon the assumption of
a Robertson-Walker metric with scalar perturbations,
plus the approximation that magnification bias and
intrinsic density fluctuations are additive.

The equations for the two observables are symmetric under the
interchange of $g\leftrightarrow \kappa$ and
$q\leftrightarrow(1+f)$.  Since $q\sim 1+f$, the lensing effects are
similar.  However the intrinsic density fluctuations $g^{\rm int}$ are
$\approx300\times$ stronger than $\kappa^{\rm int}$, breaking the
symmetry.  Density-field observations are dominated by the intrinsic
signal while convergence (shear) observations are dominated by lensing
effects.

\subsection{Degeneracies}
Equations (\ref{lenslaw})--(\ref{observables}) reveal a family of
degeneracies present in lensing observations, as described in
\citet{curvature}.  The transformations
\begin{eqnarray}
D_j & \rightarrow & D_j(1 + \alpha_0 + \alpha_1D_j + \alpha_2D_j^2) \nonumber
\\
\psi_i & \rightarrow & \psi_i(1 + \alpha_0 + 2\alpha_1D_j +
2\alpha_2D_j^2) 
\label{degen} \\
\omega_k & \rightarrow & \omega_k + 2\alpha_2 \nonumber
\end{eqnarray}
leave the observables unchanged, to first order in $\{\alpha_0,
\alpha_1D, \alpha_2D^2, \omega_kD^2\}$. It will hence be impossible
for lensing$+$density surveys to constrain $\omega_k$ or any quadratic
(in $D$) deviations to $\ln D$, unless there are prior constraints on
these variables or on $\psi, g^{\rm int},$ or $\kappa^{\rm int}$.
Constraint on these three degeneracies is unlikely to arise from
models of intrinsic clustering or alignment, since it is unlikely that
{\it a priori} models of the redshift dependence of galaxy bias could
reach high precision.  We hence expect that these degeneracies are
going to be broken by theoretical models of the potential fluctuation
power spectrum, or by other distance indicators such as supernovae or BAO.

\subsection{Limber Approximation}
To forecast the constraints on the parameters of this model, we
require a likelihood expression for the observables.  The two
fundamental assumptions we make are:
\begin{enumerate}
\item The distributions of the lensing potential, intrinsic galaxy density
  fluctuations, and 
  intrinsic shape correlations $\psi_i, g^{\rm int}_{\alpha i},$ and
  $\kappa^{\rm int}_{\alpha i}$ are described by a multivariate
  Gaussian with zero mean.
\item The Limber approximation is valid and there is no correlation
  between these variables on distinct redshift shells or between
  different spherical harmonics:
\begin{eqnarray}
\nonumber
\langle X_{i\ell m} Y_{j\ell^\prime m^\prime} \rangle & = & \delta_{ij}
\delta_{\ell \ell^\prime}
\delta_{m m^\prime}
\left(D_i^2\Delta\chi_i\right)^{-1} P_i^{XY}(\ell/D_i) \\
\label{limber}
\langle X_{i\ell m} \psi_{j\ell^\prime m^\prime} \rangle & = & -2\delta_{ij}
\delta_{\ell \ell^\prime}
\delta_{m m^\prime}
a_i\left(\ell/D_i\right)^2 P_i^{X\phi}(\ell/D_i) \\
\nonumber
\langle \psi_{i\ell m} \psi_{j\ell^\prime m^\prime} \rangle & = & \delta_{ij}
\delta_{\ell \ell^\prime}
\delta_{m m^\prime}
4a_i^2D_i^2\Delta\chi_i\left(\ell/D_i\right)^4 P_i^{\phi\phi}(\ell/D_i) 
\end{eqnarray}
where $X,Y \in \{g^{\rm int},\kappa^{\rm int}\}$, and $P_i^{XY}(k)$ is the 3-d
cross-spectrum of variables $X$ and $Y$ at epoch $a_i$.
\end{enumerate}
The first assumption insures that the likelihood of an observation is fully
specified by the expected covariance matrix of the observables.  The second
assumption implies that this covariance matrix can be expressed in
terms of the 3-d cross-power spectra of the
lensing potential, subset densities, and subset intrinsic correlations
$\{\phi_i,g^{\rm int}_{\alpha i},\kappa^{\rm int}_{\alpha i}\}$ at
each redshift shell. 

\subsection{Biases and Correlations}
A typical convention is to express the galaxy density power
$P^{gg}$ as a bias-scaled version of the mass spectrum $P^m$, plus a Poisson
shot-noise contribution, and then describing the mass-galaxy
covariance $P^{mg}$ with a correlation coefficient: 
\begin{eqnarray}
P^{gg} & = & (b^g)^2 P^m + {1 \over
  \rho} \\
P^{mg} & = & b^g r^g P^m.
\end{eqnarray}
The comoving volume number density $\rho$ of the sources determines
the shot noise for a Poisson process, but
there is no guarantee that the galaxies are distributed in the mass
distribution by a Poisson process.  Even when the galaxies do not have
Poissonian shot noise, we can still usually write the power in this
way for some bias parameter $b$; we just might keep in mind that
$r^g>1$ is formally allowed if the sources are not Poisson-distributed.

Most generally, both the
bias and correlation coefficients are different for each source {\em
  subset} as well as being functions of comoving wavenumber $k$.  Each
set $\alpha$ has a nominal redshift $z_\alpha$, and each subset has a
redshift deviation $\Delta z_{\alpha i}=z_i-z_\alpha$. Galaxies with bad
photo-z errors could easily have different bias from those with good
photo-z's; for example, highly-biased early types tend to have better
photo-z's.  So our analysis methods should allow for this
complication. 

We will adopt the bias/correlation notation for the intrinsic galaxy
density fluctuations {\em and} for the intrinsic convergence
$\kappa^{\rm int}$, except that we will parameterize the bias and
covariance with respect to the lensing potential rather than mass
distribution.  If $P^{gg}_{i\alpha\beta}$ is the 3d cross-power
between density fluctuations in subsets $\alpha i$ and $\beta i$ at
wavenumber $k$, and we
write $P^\phi_i$ for the lensing-potential 3d power spectrum, then we
express: 
\begin{eqnarray}
\nonumber
P^{gg}_{i\alpha\beta} & = &
b^g_{\alpha i}b^g_{\beta i}r^{gg}_{\alpha\beta i} 
\left({2 a_i \over 3 \omega_m}\right)^2 k^4P_i^\phi  +
{\delta_{\alpha\beta} \over \rho_{\alpha i}} \\
\label{pkk}
P^{\kappa\kappa}_{i\alpha\beta} & = &
b^\kappa_{\alpha i}b^\kappa_{\beta i}r^{\kappa\kappa}_{\alpha\beta i} 
\left({2 a_i \over 3 \omega_m}\right)^2 k^4P_i^\phi  +
{\delta_{\alpha\beta}\sigma^2_\gamma \over \rho_{\alpha i}} \\
\nonumber
P^{g\kappa}_{i\alpha\beta} & = &
b^g_{\alpha i}b^\kappa_{\beta i}r^{g\kappa}_{\alpha\beta i} 
\left({2 a_i \over 3 \omega_m}\right)^2 k^4P_i^\phi  
\end{eqnarray}
where $\rho_{\alpha i}$ is a comoving volume density of the galaxy subset
in the shell and $\sigma_\gamma$ is a measure of the shear noise per
galaxy.  These describe the normal ``shape noise'' term in the shear
power spectrum and the shot noise in the density field.  If flexions
or other observables are used to infer the convergence, then the
shape noise term may have a different form.

And if $P^{g\phi}_{i\alpha}$ is the cross-power between the density of
subset $i\alpha$ and lensing potential, we express
\begin{eqnarray}
\label{pgphi}
P^{g\phi}_{i\alpha} & = &
-b^g_{\alpha i}r^g_{\alpha i} 
{2 a_i \over 3 \omega_m} k^2P_i^\phi \\
\nonumber
P^{\kappa\phi}_{i\alpha} & = &
-b^\kappa_{\alpha i}r^\kappa_{\alpha i} 
{2 a_i \over 3 \omega_m} k^2P_i^\phi. 
\end{eqnarray}
Note that specifying the bias and correlation $b^\kappa$ and $r^\kappa$
of the intrinsic convergence with the lensing potential is equivalent
to giving the ``GI'' and
``II'' intrinsic-alignment information, in the notation of \citet{Hirata2}.

The lensing power $P^\phi$ is a function of $z$ and $k$.  The biases
and correlation coefficients $b^\kappa_{\alpha i}, r^\kappa_{\alpha i},
b^g_{\alpha i},$ and $r^g_{\alpha i}$ are functions
of $k$, the nominal redshift $z_\alpha$ of the source set, and $\Delta
z_{\alpha i}$, the difference between the subset redshift and the
nominal set redshift.

Most complicated are the cross-correlation coefficients such as
$r^{gg}_{\alpha\beta i}$, which are, most generally, functions of $k$,
$z_i$, 
and {\em both} redshift errors $\Delta z_{\alpha i}$ and $\Delta
z_{\beta i}$. In order for the covariance matrix of all these fields
to be symmetric, we require
$r^{gg}_{\alpha\alpha i}=r^{\kappa\kappa}_{\alpha\alpha i}=1$ and
the symmetry $r^{XY}_{\alpha\beta i}=r^{YX}_{\beta\alpha i}.$  
Otherwise the correlation coefficients are free to vary, subject to
the constraint that the overall correlation matrix of the potential and
all fluctuations must remain non-negative.

This parameterization of the fluctuations of the potential and the
intrinsic fluctuations is completely general---we have not introduced
any further assumptions into the model as long as all the $b$'s and
$r$'s and $P^\phi$'s are free parameters (non-negative in the last
case).

\subsection{The Two-Point Statistics}
Combining the formula for observables (\ref{observables}), the Limber formulae
(\ref{limber}), and the bias notation (\ref{pkk}), (\ref{pgphi}),
the covariance matrix for the observables $\{g_\alpha,\kappa_\alpha\}$
at a given multipole can be broken into three submatrices: 
\begin{eqnarray}
\label{gg1}
C^{gg}_{\alpha\beta} \equiv \langle g_\alpha g_\beta \rangle 
 & = & \sum_{ij} 
p_{\alpha i}p_{\beta j} \left\{
 \rule[-3ex]{0ex}{6ex} 
q_{\alpha i}q_{\beta j}\sum_n
A_{in}A_{jn}\Delta\chi_n k_n^4 P^\phi_n(k_n)
\right. \\
\nonumber
 & & \mbox{} + q_{\alpha i} A_{ij} { 2a_j  \over 3 \omega_m} D_j^{-1}
b^g_{\beta j}r^g_{\beta j}k_j^4 P^\phi(k_j) \\ \nonumber
 & &  \mbox{} + q_{\beta j} A_{ji} { 2a_i \over 3 \omega_m} D_i^{-1}
b^g_{\alpha i}r^g_{\alpha i}k_i^4 P^\phi(k_i) \\ \nonumber
 & &  \left. \mbox{} + \delta_{ij} \left({2 a_i \over 3 \omega_m}\right)^2
D_i^{-2}\Delta\chi_i^{-1}
b^g_{\alpha i}b^g_{\beta i}r^{gg}_{\alpha \beta i}k_i^4
P^\phi(k_i) 
\rule[-3ex]{0ex}{6ex} \right\} + {\delta_{\alpha\beta}\over n_\alpha}\\ 
\label{kg1}
C^{\kappa g}_{\alpha\beta} \equiv \langle \kappa_\alpha g_\beta \rangle 
 & = & \sum_{ij} 
p_{\alpha i}p_{\beta j} \left\{
 \rule[-3ex]{0ex}{6ex} 
(1+f_{\alpha i})q_{\beta j}\sum_n
A_{in}A_{jn}\Delta\chi_n k_n^4 P^\phi_n(k_n)
\right. \\
\nonumber
 & & \mbox{} + (1+f_{\alpha i}) A_{ij} { 2a_j  \over 3 \omega_m} D_j^{-1}
b^g_{\beta j}r^g_{\beta j}k_j^4 P^\phi(k_j) \\ \nonumber
 & &  \mbox{} + q_{\beta j} A_{ji} { 2a_i \over 3 \omega_m} D_i^{-1}
b^\kappa_{\alpha i}r^\kappa_{\alpha i}k_i^4 P^\phi(k_i) \\ \nonumber
 & &  \left. \mbox{} + \delta_{ij} \left({2 a_i \over 3 \omega_m}\right)^2
D_i^{-2}\Delta\chi_i^{-1}
b^\kappa_{\alpha i}b^g_{\beta i}r^{\kappa g}_{\alpha \beta i}k_i^4
P^\phi(k_i) 
\rule[-3ex]{0ex}{6ex} \right\} + {\delta_{\alpha\beta}\over n_\alpha}\\ 
\label{kk1}
C^{\kappa\kappa}_{\alpha\beta} \equiv \langle \kappa_\alpha \kappa_\beta \rangle 
 & = & \sum_{ij} 
p_{\alpha i}p_{\beta j} \left\{
 \rule[-3ex]{0ex}{6ex} 
(1+f_{\alpha i})(1+f_{\beta j})\sum_n
A_{in}A_{jn}\Delta\chi_n k_n^4 P^\phi_n(k_n)
\right. \\
\nonumber
 & & \mbox{} + (1+f_{\alpha i}) A_{ij} { 2a_j  \over 3 \omega_m} D_j^{-1}
b^\kappa_{\beta j}r^\kappa_{\beta j}k_j^4 P^\phi(k_j) \\ \nonumber
 & &  \mbox{} + (1+f_{\beta j}) A_{ji} { 2a_i \over 3 \omega_m} D_i^{-1}
b^\kappa_{\alpha i}r^\kappa_{\alpha i}k_i^4 P^\phi(k_i) \\ \nonumber
 & &  \left. \mbox{} + \delta_{ij} \left({2 a_i \over 3 \omega_m}\right)^2
D_i^{-2}\Delta\chi_i^{-1}
b^\kappa_{\alpha i}b^\kappa_{\beta i}r^{\kappa\kappa}_{\alpha \beta i}k_i^4
P^\phi(k_i) 
\rule[-3ex]{0ex}{6ex} \right\} +
\delta_{\alpha\beta}\left({\sigma^2_\gamma\over n}\right)_\alpha
\end{eqnarray}
The comoving wavevector is $k_i=\ell/D_i$.
In each equation, note that only one of the last three terms is
non-zero, depending on whether $j<i$, $j>i$, or $j=i$, respectively.
For the shear-shear correlation $C^{\kappa\kappa}$,
the $i=j$ term is recognizable as the ``II''
intrinsic-correlation effect of \citet{Hirata2}, while the $i<j$
and $j>i$ terms are their ``GI'' effect.  

Note that
the last term in the density-shear expression $C^{\kappa g}$
is an additional intrinsic-correlation term, between the galaxy
density and the intrinsic shapes, which is distinct from the
covariance between the lensing potential and shear.  This galaxy-shear
correlation has been constrained in the context of systematic errors
to ``galaxy-galaxy'' lensing, \eg
\citet{BernsteinNorberg,FLMvdBYJPM,Hetal04}. 

Examining the density-density correlation $C^{gg}$ we find the final
term has the normal expected form, but the first three terms describe
correlations induced by lensing magnification. 

Finally we note that the covariance matrix manifests the same
symmetries for $g\leftrightarrow\kappa$ that were discussed at the end
of \S\ref{observsec}.

\subsection{Likelihood and Fisher matrix}
Under our Gaussian assumption, the likelihood functions for the
observables are independent at each multipole $\ell m$.  We define a
data vector
${\bf d}_{\ell m}$ to be the union of the $g_\alpha$ and
$\kappa_\alpha$ observables at each multipole, and ${\bf C}_{\ell}$ to
be the covariance matrix derived above.  Under our Gaussian
assumption, the total likelihood for the survey is
\begin{equation}
\label{likelihood}
-2\ln L = \sum_{\ell m} \left[ {\bf d}_{\ell m}^T {\bf C}_\ell^{-1} {\bf
  d}_{\ell m} + \ln | {\bf C}_\ell | \right].
\end{equation}
Forecasts of survey performance are made using the Fisher matrix.  The
usual formula for zero-mean Gaussian distributions applies
\citep{TTH97}.  We reduce the mode sum to a series of $N_\ell$ bins centered
on multipoles $\ell_i$, then the Fisher matrix element for parameters
$p$ and $q$ is
\begin{equation}
\label{fisher}
F_{pq} = \sum_{i=1}^{N_\ell} {(2\ell_i+1)\Delta \ell_i\,f_{\rm
    sky}\over 2}\,
{\rm Tr} \left[{\bf C}^{-1}_{\ell_i} {\partial {\bf C}_{\ell_i} \over \partial p}
{\bf C}^{-1}_{\ell_i} {\partial {\bf C}_{\ell_i} \over \partial q}\right].
\end{equation}
Examination of Equations~(\ref{gg1})--(\ref{kk1}) shows that 
all derivatives of ${\bf C}$ with respect to parameters are very simple.
The calculation of the Fisher matrix is reduced to rapid linear
algebra, significantly accelerated by exploiting the very sparse
nature of most of the derivative matrices.

We have thus succeeded in producing a likelihood function for the most
general joint lensing$+$density survey, for the case of Gaussian
likelihoods limited to 2-point statistics.  Given a likelihood we can
of course form a Fisher matrix for forecasting, or we can execute a
maximum-likelihood analysis of real data.  Since this likelihood
function makes no mention of a particular dark-energy theory, we see
that the parameterization chosen here permits a highly flexible
analysis.  Indeed no theory of gravity or initial conditions of the
Universe have been assumed either, just the existence of a Newtonian
gauge metric on a RW background cosmology.  The lensing-potential
power spectrum $P^\phi(k,z)$ appears as a series of free parameters,
as do the bias and correlation coefficients of the galaxy density and
intrinsic alignments.

We have variables
that describe, in the most general possible fashion, the important
systematic errors, excepting additive shear contamination:  
\begin{enumerate}
\item Uncertainty in power-spectrum theory will be expressed through
  prior distributions on the $P^\phi_i$ parameters. 
\item Shear calibration errors arises through finite prior uncertainty
  on the $f_{\alpha i}$. 
\item Magnification-bias calibration errors arises through finite prior
  uncertainty on the $q_{\alpha i}$. 
\item Intrinsic alignments are embodied through the $b^\kappa$,
  $r^\kappa$, $r^{g\kappa}$, and $r^{\kappa\kappa}$ coefficients. 
\item Redshift-distribution errors are manifested through the
  uncertainties in the $p_{\alpha i}$ probabilities. 
\end{enumerate}

The cost of this great generality is that there are a huge number of
nuisance parameters, enough to make us doubt whether the
maximum-likelihood analysis---or even the Fisher-matrix analysis!---is
feasible.  

\subsection{Parameter Inventory}
The WL survey covariance matrix has a
horrendously large number of parameters.  The cosmological treasure
lies in these:
\begin{itemize}
\item The 2 global cosmological parameters $\omega_m$ and $\omega_k$.
\item The distances $D_i$, which encode the expansion history of the
  Universe in $N_z$ steps.  The $\Delta\chi_i$ and the Hubble
  parameters  $h(z_i)$ can be expressed in terms of these and $\omega_k$.
\item The metric-potential power spectra $P_i^\phi$, which describe
  the growth of 
  dark-matter structure.  For $N_\ell$ bins in $\ell$, there will be
  $N_\ell N_z$ distinct matter-power parameters in the model.  A
  prediction for the growth of potential fluctuations will typically be an
  important element of any cosmological scenario under test, so the
  $P^\phi_i$ can be replaced as parameters by a much smaller number of
  cosmological parameters.
\end{itemize}
There are then a large number of nuisance parameters.  If there are
$N_{ss}$ non-empty source subsets, the nuisance parameters are 
\begin{itemize}
\item The redshift-distribution parameters $p_{\alpha i}$, with
  $N_{ss}-N_s$ degrees of freedom.
\item The shear-calibration errors $f_{\alpha i}$, another $N_{ss}$
  degrees of freedom.
\item The magnification-bias coefficients $q_{\alpha i}$, another $N_{ss}$
  degrees of freedom.
\item The source-density biases $b^g_{\alpha i}$ and correlation coefficients
  $r^g_{\alpha i}$ with respect to $\phi$, which may be scale-dependent,
  yielding $2N_\ell N_{ss}$ degrees of freedom.
\item The intrinsic-alignment power and correlations with the mass,
  $b^\kappa_{\alpha i}$ and $r^\kappa_{\alpha i}$, another $2N_\ell
  N_{ss}$ parameters. 
\item The correlation coefficients $r^{gg}_{\alpha\beta i}$, 
$r^{\kappa g}_{\alpha\beta i}$, and $r^{\kappa\kappa}_{\alpha\beta
  i}$, which also may be scale-dependent.  The number of such
parameters is $\approx 3N_\ell N_{ss}^2/2N_z$.
\end{itemize}
The number of nuisance parameters for a non-parametric analysis is
enormous.  If we are analyzing a photo-z survey with typical errors
$\Delta z\approx 0.05(1+z)$, then we would typically want to space the
redshift shells logarithmically in $1+z$, with $\Delta\ln a \approx
0.02$ so that we resolve the redshift distribution of each photo-z
bin.  In this case, $N_z\approx100$ bins span $0<z<5$, and
we will require $N_{ss}\gtrsim 1000$ if we track all subsets out to
$\pm3\sigma$ of the photo-z distribution. 

To reduce the dimensionality of the likelihood function, we can
replace many of the discrete nuisance parameters by the values of
parameterized functions for the nuisance variables.
Table~\ref{nuisancefunc} lists the variables in the WL likelihood
function that can be replaced by parametric functions. The nuisance
variables are functions of: wavevector $k$; redshift $z$; and redshift
difference $\Delta z = z_\alpha - z_i$ between the nominal and true
redshifts of a source subset.
In later sections we will describe the parametric functions
that we have implemented to reduce the number
of degrees of freedom in the model.  Each time we introduce a
parametric function, we need to choose a fiducial parameter set and a
prior distribution for the parameters.

\begin{deluxetable}{lcc}
\tablewidth{0pt}
\tablecaption{Nuisance variables that can be replaced by functions}
\tablehead{
\colhead{Description} &
\colhead{Discrete variables} &
\colhead{Parametric function}
}
\startdata
Lensing potential power spectrum & $P^\phi_i$ & $P^\phi(k,z)$ \\
Shear calibration error & $f_{\alpha i}$ & $f(z, \Delta z)$ \\
Magnification bias & $q_{\alpha i}$ & $q(z, \Delta z)$ \\
Redshift distribution & $p_{\alpha i}$ & $p(z, \Delta z)$ \\
Source density bias & $ b^g_{\alpha i}$ & $b^g(k, z, \Delta z)$ \\
Density-mass correlation & $ r^g_{\alpha i}$ & $r^g(k, z, \Delta z)$ \\
Intrinsic alignment bias & $ b^\kappa_{\alpha i}$ & $b^\kappa(k, z, \Delta z)$ \\
IA-density correlation & $ r^\kappa_{\alpha i}$ & $r^\kappa(k, z, \Delta z)$ \\
Density-density x-correlation & $r^{gg}_{\alpha\beta i}$ &
$r^{gg}(k,z,\Delta z_\alpha, \Delta z_\beta)$ \\
Density-IA x-correlation & $r^{g\kappa}_{\alpha\beta i}$ &
$r^{g\kappa}(k,z,\Delta z_\alpha, \Delta z_\beta)$ \\
IA-IA x-correlation & $r^{\kappa\kappa}_{\alpha\beta i}$ &
$r^{\kappa\kappa}(k,z,\Delta z_\alpha, \Delta z_\beta)$
\enddata
\label{nuisancefunc}
\end{deluxetable}

\section{Spectroscopic Redshift Likelihood}
\label{speclike}
If we draw a single
member from source set $\alpha$ and measure its spectroscopic redshift in
an unbiased 
fashion, then by definition the likelihood of the spectroscopic
redshift being on shell $i$ is $p_{\alpha i}.$  If we measure
$N_\alpha^{\rm spec}$ redshifts, and find that $N^{\rm spec}_{\alpha
  i}$ are on shell $i$, 
then the likelihood is
\begin{equation}
\ln L = \sum_i N^{\rm spec}_{\alpha i} \ln p_{\alpha i}.
\end{equation}
This is true if the redshifts are statistically independent, which
requires that they be dispersed across the sky to eliminate source
correlations.  We assume this limit.

Following \citet{MaBernstein}, the
Fisher matrix for the parameters $\{p_{\alpha i}\}$ resulting from the
unbiased spectroscopic observations is
\begin{equation}
F_{\alpha i \beta j} = \left\langle {\partial^2 (-\ln L) \over \partial p_{\alpha i}
\partial p_{\alpha j} } \right\rangle = N^{\rm spec}_\alpha {\delta_{ij} \over
p_{\alpha i}}.
\label{specfish}
\end{equation}
We add this Fisher information to the density-lensing Fisher matrix
(\ref{fisher}) 
when considering the constraints offered by a WL survey that is
combined with an {\em unbiased} spectroscopic redshift survey drawn
from one or more of the source population sets.

We do not in general presume any functional form for the $p_{\alpha
  i}$ redshift distributions when the sets are assigned from
photo-z's.  We adopt fiducial values either from an analytic form or from
a simulation of photo-z performance.  Then we leave all the $p_{\alpha
  i}$ as free parameters in the Fisher matrix, adding the
spectroscopic-survey Fisher matrix (\ref{specfish}) if appropriate to
the planned 
experiment.  Note that the cross-correlations in the WL survey data
offer constraints on the redshift distribution even if there is no
unbiased spectroscopic survey ($N^{\rm spec}_\alpha=0$).

\section{An Implementation}
A package of {\tt C++} classes implements the Fisher matrix
calculation for Gaussian lensing$+$density observations, the
spectroscopic-survey Fisher matrix, plus the
models for lensing-potential power and nuisance functions
described below.  From these classes we can construct numerous
applications, the most obvious being a Fisher-matrix forecast of
cosmological constraints from the combination of a photometric galaxy
lensing/density survey, plus a redshift survey to constrain the
photo-z distribution.
We list in Table~\ref{params1} input fields for this forecasting
implementation.  Further program inputs
are listed in later sections which detail the models for the lensing
power spectrum and nuisance parameters that we describe below and
adopt for this implementation.

In the current implementation we assume the source galaxies to be
binned solely by photo-z, but generalizations are possible, \eg including a
second population of source galaxies
sets that are observed spectroscopically.  Note that when additional
galaxy populations are introduced, we need a new set of nuisance functions
to describe them.  Furthermore, we need to model the
cross-correlations between all galaxy populations.

The calculation of the Fisher matrix takes $<1$ minute per
multipole bin using a single core of a typical current-epoch desktop
CPU.  Total execution time for a forecast is 10--20 minutes with the
default parameters, with the most time-consuming operation being the
marginalization over bias model parameters.  The execution time is
very sensitive to the redshift-shell width $\Delta\ln a$ and to the
width of the photo-z error distribution, as these control the number
of subsets and the parameter count.

\begin{deluxetable}{cll}
\tablewidth{0pt}
\tablecaption{Controlling inputs to Fisher forecast}
\tablehead{
\colhead{Parameter name} &
\colhead{Description} &
\colhead{Default}
}
\startdata
{\tt fsky} & $f_{\rm sky}$, imaging sky coverage & 0.5 \\
{\tt minLogL} & $\log_{10} \ell_{\rm min}$, minimum multipole & 1.0 \\
{\tt maxLogL} & $\log_{10} \ell_{\rm max}$, maximum multipole & 3.5 \\
{\tt logLStep} & $\Delta \log_{10} \ell$, multipole bin width in dex
& 0.3 \\
{\tt zmax} & $z_{\rm max}$, redshift of most distant shell & 3.5 \\
{\tt dlna} & $\Delta \ln a$, width of distance shells & 0.03 \\
{\tt coreDLna} & Maximum $|\ln (1+z_i)/(1+z_\alpha)|$ of subsets &
0.15 \\
{\tt sigGamma} & $\sigma_\gamma$, shape noise per source galaxy & 0.24
\\
{\tt zdist} & String specifying fiducial source $n_\alpha$ and
$p_{\alpha i}$ & \nodata \\
{\tt logNSpec} & $\log_{10} N_{\rm spec}$ & 5. \\
{\tt outfile} & Root name for output files & \nodata 
\enddata
\label{params1}
\end{deluxetable}

\section{Power Spectrum Models}
\label{powermodel}
In most cosmological models, theory will offer strong guidance to the
form of the lensing-potential power spectrum $P^\phi(k,z)$.
In most forecasting or data-reduction codes this is a
fully deterministic function of a small number of cosmological
parameters.  In our analysis, however, the theoretical prediction is
taken as the mean $P^\phi$ value of a prior distribution of finite
uncertainty. 

\subsection{Central Model}
The WL likelihood given above can be calculated for any model that
predicts $P^\phi$.  We have chosen to implement a model that allows
for failure of General Relativity in describing growth of structure;
but other models are possible if one wishes to test the Poisson
equation or other tenets of General Relativity.  Under the conditions
\begin{enumerate}
\item The potential and the mass-energy density are related by the
  Poisson equation of General Relativity, and
\item Non-relativistic matter is the only significant inhomogeneous
  component of the 
  Universe, \ie there is no dark-energy clustering, and
\item Matter is conserved, $\bar \rho_m \propto a^{-3}$, and
\item $\Phi = -\Psi$, as in the absence of anisotropic stress for
  General Relativity
\end{enumerate}
then the potential power spectrum is related to the matter-density
fluctuation spectrum $P^m$ via
\begin{equation}
k^4P^\phi(k,a) = \left({3\omega_m \over 2a}\right)^2 P^m(k,a).
\label{poisson}
\end{equation}
Under these conditions, the linearized perturbations to the metric
grow in a scale-free manner so we can write
\begin{equation} 
P^\phi_{\rm lin}(k, a) = g_\phi^2(a) P^\phi_{\rm prim}(k) T^2(k).
\end{equation}
In our current code, the primordial power spectrum is a power law 
\begin{equation}
\Delta^2_{\rm prim}(k) \equiv {k^3 \over 2 \pi^2}P^\phi_{\rm prim}(k)
 = \left({3 \Delta_\zeta \over 5}\right)^2 (k/k_0)^{n_s-1}.
\end{equation}
The curvature variation $\Delta_\zeta$ and spectral index $n_s$ are
free parameters.
A running of the slope could easily be
added. The normalization wavenumber $k_0$ must be set by some
convention.  We typically adopt the 5-year WMAP parameters as fiducial
values \citet{WMAP5}.

The transfer function $T(k)$ is taken from \citet{EH}.  It is a function of
the matter and baryon densities $\omega_m$ and $\omega_b$.  The impact
of massive neutrinos could be added to the transfer function if
desired.  We ignore the baryon acoustic oscillations; experiments that
try to exploit them will generate a distinct Fisher matrix for them.

The map from $P^\phi_{\rm lin}$ to the nonlinear $P^\phi_{\rm nl}$ is derived
using the prescription of \citet{Smith} for nonlinear $P^m$, combined
with the Poisson equation (\ref{poisson}).  The
Smith {\it et al.} formula also requires knowledge of $\Omega_m$ at
the desired epoch, but it can be expressed in terms of other
quantities that are already in our model: $\Omega_m(z_i)=\omega_m
a_i^{-3} h^{-2}(z_i).$  We do not expect the \citet{Smith} formula to
describe non-linear growth to high accuracy for all (or any) cosmologies.
It does however capture the dependence of non-linear
power on cosmological parameters to a level that suffices for
forecasting purposes.

In General Relativity, the growth function $g_\phi(a)$ is determined by the
expansion history $H(z)$.  Defining $F=\ln (ag_\phi)$, the growth
equation is
\begin{equation}
\label{growtheq}
F^{\prime\prime} + \left( F^\prime\right)^2 + F^\prime\left(2+ {d\ln h
    \over d \ln a}\right) = {3 \omega_m \over 2 h^2 a^3},
\end{equation}
where a prime denotes differentiation with respect to $\ln a$.  Note
that $F^\prime$ is the quantity $d\ln g_m/d\ln a$ that appears in the
peculiar-velocity power spectrum for a tracer of mass.  

If GR holds, then the above relations fully specify the model for
$P^\phi$ given $\{\omega_m, \omega_b, n_s, \ln \Delta_\zeta\}$ plus
the expansion history, which in turn is given by $\{D_i,
\omega_k\}$. As a test of GR, we allow the growth function arbitrary
deviations from the $F_{\rm fid}$ that solves the GR growth equation for
the fiducial expansion history:
\begin{equation}
\ln ag_\phi(a_i) = F_{\rm fid}(a_i) + \delta F_i.
\end{equation}
The $\delta F_i$ become parameters of the likelihood function.

\subsection{Model Errors}
An important WL systematic is the expected finite accuracy in
  theoretical modeling of the power spectrum.  We hence introduce
  an error function to describe the describe the (logarithmic)
  difference between the power $P^\phi$ and the value predicted by the
  parametric model described in the previous paragraphs:
\begin{equation}
\ln P^\phi(k,a) = \ln P^\phi_{\rm nl}(k,a) + \delta \ln P(k,a).
\end{equation}
The nuisance function $\delta\ln P$ will be described with the
``$kz$'' parametric form described in Appendix~\ref{functions}.
We parameterize the $\delta \ln P$ function by its values $\delta
P_{ij}$ at a grid of points $(k_i, a_j)$ regularly spaced in $\ln k$
and $\ln a$.  The $\delta\ln P$ is linearly interpolated between grid
points. 
The $\delta P_{ij}$ become free parameters of
the model and hence
parameters in the likelihood function. We then place an independent
Gaussian prior on each $\delta P_{ij}$ which has mean of zero and a
standard deviation of
\begin{eqnarray}
\sqrt{{\rm Var}(\delta P_{ij})} & = & 0.012 f_{\rm Zhan} \cases {
1 + 5\log_{10}(k_i/k_1) & $k_i>k_1$ \cr
(k_i/k_1)^{1+a_j} & $k_i<k_1$ }  \\
k_1 & \equiv & a^{-2.6} \,{\rm Mpc}^{-1}.
\end{eqnarray}
This function is a fit to an estimate, supplied by Hu Zhan, of the
impact of baryonic physics on the mass power spectrum \citet{ZhanKnox,Jing}.
We scale the overall size of the theory-error systematic with the
control scalar $f_{\rm Zhan}$.  We can also adjust the density $\Delta
\ln k$ and $\Delta \ln a$ at which the $\delta P_{ij}$ grid points are
spaced. This corresponds to setting some coherence length for theory
errors in this space.  In \S\ref{tuneup} we will investigate the choice of
these grid spacings.

The procedure above means that we replace the $P^\phi_i$ as parameters in
our likelihood with a new (and hopefully smaller) set:
\begin{itemize}
\item The small set $\{\omega_m, \omega_b, \Delta^2_\zeta, n_s\}$ that
  control the linear power spectrum.
\item The $\{\delta F_i\}$ which define the growth function vs redshift.
\item A grid of theory error values $\delta P$, which are nuisance
  parameters to marginalize after construction of a Fisher matrix or
  likelihood.  We have physically-based priors to apply to these
  before marginalization. 
\end{itemize}
Table~\ref{params2} lists the input fields for the part of our
forecasting code which constructs the power-spectrum model.  The WMAP5
$\Lambda$CDM cosmology provides the fiducial values of all
lensing power values; the program inputs define the behavior of the
deviations from the theoretical model and the prior expectations on
the size of such deviations.
\begin{deluxetable}{cll}
\tablewidth{0pt}
\tablecaption{Power spectrum inputs to Fisher forecast}
\tablehead{
\colhead{Parameter name} &
\colhead{Description} &
\colhead{Default}
}
\startdata
{\tt zhan} & $f_{\rm Zhan}$, power-spectrum theory uncertainty
relative to baryonic effects & 0.5 \\
{\tt psDlnk} & $\Delta \ln k$, node spacing of power-spectrum theory
errors in $k$ & 1.0 \\
{\tt psDlna} & $\Delta \ln a$, node spacing of power-spectrum theory
errors in $a$ & 0.5  
\enddata
\label{params2}
\end{deluxetable}

\section{Non-parametric nuisance modeling}
\label{nuisance}
\subsection{General comments on nuisance functions}
The WL likelihood contains many nuisance parameters that are discretized
representations of nuisance {\em functions}.  It is common in the
literature to assign some parametric form to a nuisance
function, then marginalize over the parameters of the nuisance
function to recover a purely cosmological likelihood.  This can be a
very dangerous approach: if the nuisance function does {\em not}
in actuality follow the assumed form, then the process is invalid and
we may have greatly overestimated the power of the experiment to
remove the systematic error from the signal.  When marginalizing over
a systematic, we must be sure that the assumed parametric form is
sufficiently flexible to include any expected manifestation of the
systematic.  For example we should not assume that systematics scale
linearly with redshift unless there is a physical reason to expect
this.

It is unfortunately not possible to model a completely free function
with a finite number of free parameters.  This becomes possible,
however, if we limit the bandwidth of variation in the function.
As an example consider our power-spectrum theory error function
$\delta \ln P(k,z)$.  We could decompose $\delta \ln P$ into
Fourier modes or polynomial terms over its finite $(k,z)$ domain.
Retaining a finite number of modes or terms leads to a tractable
parameterization, albeit with a maximum frequency or polynomial order
that defines a coherence length for the reconstructed function. For
$\delta \ln P$ we choose to limit the bandwidth using linear interpolation
between a 2d grid of specified values.
In
Appendix~\ref{functions} we describe the family of functions that we
use to model the nuisance functions of $(k, z, \Delta z)$ that are
common in the WL likelihood analysis ({\it cf.}
Table~\ref{nuisancefunc}).\footnote{
In practice we use $\ln a$ and $\Delta\ln a$ to specify each subset's
nominal redshift and redshift error, but in the text we will stick
with $(z,\Delta z)$ to reduce the clutter.}
  The Appendix describes both the
functional form, and the prior likelihoods on the parameters that are
used to give the nuisance function the desired $RMS$ uncertainty.

These models of nuisance functions are non-parametric in the sense of
being able to reproduce very general types of behavior once the
bandwidth is specified.  The question remains: what is the proper
choice of bandwidth to allow the nuisance function?  Our approach is
to find the bandwidth which {\em causes the most damage to
  cosmological constraints} under a prior that specifies the expected RMS
fluctuations in the nuisance function.  This is the most conservative
approach.  Typically one finds the following: if the nuisance function
is given a highly coherent, low-order functional form, then it is
easily distinguished from cosmological signals and can be marginalized
away with little damage to cosmological constraints.  On the other
hand if the nuisance-function bandwidth is very high, then the
broad WL kernel tends to average away the nuisance signal, leaving 
little trace in the cosmology.  There is an intermediate point where
the systematic error is most easily confused with cosmology.  The
conservative approach is to find this regime and use it for modeling
the systematic error.  In \S\ref{tuneup} we will find coherence lengths in $z$
and $k$
at which our systematics are most damaging.

\subsection{Redshift Distributions}
We specify the fiducial values of the photo-z distribution $n_\alpha$
and error probabilities $p_{\alpha i}$ either with analytic formulae
(\eg photo-z errors Gaussian in $\ln a$), or by taking the output of
a simulation of galaxy detection and photo-z assignment for the chosen
survey.  For spectroscopic samples or the CMB source plane there are
no photo-z errors at all.

We do not place any parametric form or prior assumption on the
$p_{\alpha i}$.  All redshift constraints arise either
from the lensing survey data itself or from additional spectroscopic
data.  The likelihood arising from spectroscopic redshift samples is
described in \S\ref{speclike}.

\subsection{Shear calibration and magnification bias}
The shear calibration factors $f_{\alpha i}$ and magnification bias
coefficients $q_{\alpha i}$ are, most generally, distinct in every
subset.  We use the ``$z\Delta z$'' functional
form described in Appendix~\ref{functions} to generate the $f_{\alpha
  i}$ and $q_{\alpha i}$ from a smaller set of function parameters.
Each function is specified by a polynomial function of $\Delta
z$; polynomial coefficients are interpolated between grid points
equally spaced in $\ln a$ at intervals $\Delta\ln a$.

We set the fiducial functions to be
$f_{\alpha i}=0,$ and $q_{\alpha i}=q_{\rm fid}$ independent of
$(k,z)$.  The priors on the
polynomial coefficients are chosen to yield a chosen RMS variation of
$f$ or of $q$.  As detailed in Appendix~\ref{functions}, we also
specify whether the nuisance function varies mostly along the $z$
direction or along the $\Delta z$ direction of its domain.

Table~\ref{params3} lists the program inputs necessary to specify the
model for $f$ or $q$: their functional form, fiducial values, and
priors on deviations from the fiducial.

\begin{deluxetable}{cll}
\tablewidth{0pt}
\tablecaption{Shear calibration and magnification bias inputs to Fisher forecast}
\tablehead{
\colhead{Parameter name} &
\colhead{Description} &
\colhead{Default}
}
\startdata
{\tt fRMS} & $f_{\rm RMS}$, RMS variation of $f$ allowed under prior &
0.01 \\
{\tt qFid} & $q_{\rm fid}$, fiducial value for all $q_{\alpha i}$ & 1.0 \\
{\tt qRMS} & $q_{\rm RMS}$, RMS variation of $q$ allowed under prior &
0.1 \\
{\tt fqDlna} & $\Delta\ln a$, node spacing for $f$ \& $q$ models & 0.5
\\
{\tt fqDzOrder} & Order of polynomial used to model $\Delta z$
dependence of $f,q$
& 2 \\
{\tt fqVarFracDZ} & Fraction of $f,q$ variance that due to $\Delta z$
dependence & 0.5  
\enddata
\label{params3}
\end{deluxetable}

\subsection{Galaxy correlation coefficients}
The correlation coefficient
$r^g_{\alpha i}$ is, most generally, different at each subset
and at each multipole $\ell$.  We model $r^g$ using the 
``$kz\Delta z$'' function form described in Appendix~\ref{functions}.
These functions are polynomial in $\Delta z$, with the polynomial
coefficients linearly interpolated
from a grid in $(\ln k, \ln a)$ space.  This grid of
polynomial coefficients replaces the $r^g_{\alpha i}$
as parameters in the likelihood.

For the fiducial correlation
coefficient, we interpolate smoothly between a linear and non-linear
limit according to the value of $\Delta^2_{\rm lin}(k,z)=k^3P^m_{\rm
  lin}(k,z)/2\pi^2$:
\begin{equation}
r^g_{\rm fid}(k,z) = {r^g_{\rm NL} + r^g_{\rm L} \over 2}
 + { (r^g_{\rm NL} - r^g_{\rm L})  \over \pi} \tan^{-1} \left[
{ \ln \Delta_{\rm lin}(k,z) \over W} \right].
\label{lnlinterp}
\end{equation}
The constant $W$ sets the width of the transition from the linear to
nonlinear regime.  We set $W=1$ unless otherwise noted.

Each polynomial coefficient at each grid point is assigned an
independent Gaussian prior.  These are selected to yield a preselected
RMS variation $r^g_{\rm RMS}$. The RMS prior
uncertainties are interpolated in $(k,z)$ space
between linear and 
non-linear limiting values $r^g_{\rm RMS,L}$ and $r^g_{\rm RMS,NL}$
using the same functional form (\ref{lnlinterp}).

Table~\ref{params4} lists the program inputs needed to specify the
galaxy bias and correlation models.

\subsection{Galaxy bias}
We expect $b^g$ to vary quite strongly with $z$, as the more distant
source galaxies are likely intrinsically very bright and highly
biased. There may also be a strong dependence of $b^g$ on $\Delta z$
because both $\Delta z$ and the bias may couple strongly to galaxy
spectral type.  Variation with $k$ should be weaker.  We hence define $b^g$
to be the sum of two functions,
\begin{equation}
b^g = b^g_{\rm coarse}(z,\Delta z) + b^g_{\rm fine}(k,z,\Delta z).
\end{equation}
The fiducial values are $b^g_{\rm coarse}=b^g_{\rm fid}, b^g_{\rm
  fine}=0$.  

The coarse
contribution is given a very weak prior, but can only vary slowly with
$z$: $\Delta\ln a=0.5$ by default for the $b^g_{\rm coarse}$ grid nodes.

The $b^g_{\rm fine}$ function is interpolated between the same $(\ln
k, \ln a)$ grid points as the correlation coefficient $r^g$. The prior
on each $b^g_{\rm fine}$ node is arranged to give RMS uncertainty
that is interpolated between linear and nonlinear limits $b^g_{\rm
  RMS,L}$ and $b^g_{\rm RMS,NL}$ just as for $r^g$.

\begin{deluxetable}{cll}
\tablewidth{0pt}
\tablecaption{Galaxy bias inputs to Fisher forecast}
\tablehead{
\colhead{Parameter name} &
\colhead{Description} &
\colhead{Default}
}
\startdata
{\tt bg} & $b^g_{\rm fid}$, fiducial galaxy bias & 1.5 \\
{\tt rgL} & $r^g_{\rm L}$, fiducial galaxy correlation coeff at linear
limit & 0.9 \\
{\tt rgNL} & $r^g_{\rm NL}$, fiducial galaxy correlation coeff at non-linear
limit & 0.6 \\
{\tt biasDlnk} & $\Delta\ln k$ interpolation grid step for bias
models & 1.0 \\
{\tt biasDlna} & $\Delta\ln a$ interpolation grid step for bias
models & 0.1 \\
{\tt brgRMSL} & $b^g_{\rm RMS,L}$ and $r^g_{\rm RMS,L}$ RMS prior
variation for bias and correlation, linear limit & 0.05 \\
{\tt brgRMSNL} & $b^g_{\rm RMS,NL}$ and $r^g_{\rm RMS,NL}$ RMS prior
variation non-linear limit & 0.10 \\
{\tt brVarFracDZ} & Fraction of $b^g,r^g$ variance due to $\Delta
z$ dependence & 0.2 \\
{\tt bgCoarseDlna} & $\Delta\ln a$ node grid spacing for $b^g_{\rm coarse}$
 & 0.5 \\
{\tt bgCoarseZRMS} & RMS prior variation of $b^g_{\rm coarse}$ at
$\Delta z=0$ & 0.5 \\
{\tt bgCoarseDZRMS} & RMS prior variation of $b^g_{\rm coarse}$ at
fixed $z$ & 0.3 \\
{\tt kgg} & $K^{gg}_{\rm fid}$, fiducial value of $K^{gg}(k,z)$
cross-corr spec & 2. \\
{\tt kggRMS} & $K^{gg}_{\rm RMS}$, prior variation on $K^{gg}(k,z)$ &
1.
\enddata
\label{params4}
\end{deluxetable}

\subsection{Intrinsic Alignments}
The strength of intrinsic alignments are specified by the
$b^\kappa_{\alpha i}$ and $r^\kappa_{\alpha i}$ values at each
multipole $\ell$.  As for the galaxy density, the free-parameter count
can be reduced by specifying parametric functions $b^\kappa,
r^\kappa$ of $(k, z, \Delta 
z)$ instead.  Each of these two functions modeled using the
``$kz\Delta z$'' form described in Appendix~\ref{functions}, just as
for $r^g$.  The fiducial $b^\kappa_{\rm fid}$ and 
$r^\kappa_{\rm fid}$ are
taken as constant over the entire domain.  The RMS variation
$b^\kappa_{\rm RMS}$ and $r^\kappa_{\rm RMS}$ in the
priors of these functions are also taken to be constant over the domain.
We take $b^\kappa_{\rm RMS}=|b^\kappa_{\rm fid}|$, because we
expect that the best constraints on IA to always arise from self-calibration
of WL surveys rather than through any external modeling or
prior.  Roughly speaking, the IA measured by \citet{Mandelbaum06} for
the SDSS population corresponds to $b^\kappa \approx -0.003$
\citep{BridleKing}, which we will normally take as our fiducial model
for IA.  Setting $b^\kappa_{\rm fid}=0$ turns off the IA systematic entirely.

Table~\ref{params5} lists the program inputs needed to specify the
functional form for intrinsic alignments, the fiducial values, and the
prior constraints.  Note that we assume the intrinsic-alignment
functions to be defined on the same $(\ln k, \ln a)$ grid as the
galaxy bias and covariance functions.

\begin{deluxetable}{cll}
\tablewidth{0pt}
\tablecaption{Intrinsic-alignment (IA) inputs to Fisher forecast}
\tablehead{
\colhead{Parameter name} &
\colhead{Description} &
\colhead{Default}
}
\startdata
{\tt bk} & $b^\kappa_{\rm fid}$, fiducial intrinsic alignment &
$-0.003$ \\
{\tt rk} & $r^\kappa_{\rm fid}$, fiducial correlation between mass and
intrinsic alignment & 0.7 \\
{\tt rkRMS} & $r^\kappa_{\rm RMS}$, RMS prior variation in $r^\kappa$ & 0.2 \\
{\tt kkk} & $K^{\kappa\kappa}_{\rm fid}$, fiducial value of
$K^{\kappa\kappa}(k,z)$ & 1. \\
{\tt kkkRMS} & $K^{\kappa\kappa}_{\rm RMS}$, RMS prior variance of $K^{\kappa\kappa}$
 & 1.\\
{\tt skg} & $s^{g\kappa}_{\rm fid}$, fiducial IA-bias
cross-correlation & 0. \\
{\tt skgRMS} & $s^{g\kappa}_{\rm RMS}$, RMS prior on IA-bias
cross-correlation & 0.3 
\enddata
\label{params5}
\end{deluxetable}

\subsection{Cross-correlation coefficients}
The cross-correlation coefficients $r^{gg}_{\alpha\beta i},
r^{g\kappa}_{\alpha\beta i}$, and $r^{\kappa\kappa}_{\alpha\beta i}$ are even
more complex because each depends on $k, z$, plus two subsets' $\Delta
z_{\alpha i}$ and $\Delta z_{\beta i}$.  We find it infeasible to
construct nuisance-function templates spanning 4 dimensions.  We
therefore simplify by first writing 
\begin{equation}
r^{g\kappa}_{\alpha \beta i}  =  r^g_{\alpha i} r^\kappa_{\beta i} 
 + s^{g\kappa}_{\alpha \beta i} 
\sqrt{\left[1-(r^g_{\alpha i})^2\right]
      \left[1-(r^\kappa_{\beta i})^2\right]}.
\end{equation}
A value $|s^{g\kappa}_{\alpha \beta i}|\le 1$ is necessary (but not
sufficient) to keep the
mass-galaxy covariance 
matrix from acquiring non-physical negative eigenvalues.  In principle
the functional form of $s^{g\kappa}$ must vary over four
dimensions, but we make the gross simplification that it is constant
for the survey, since we expect this type of cross-correlation to have
minimal effect on cosmological constraints.  The program thus requires
simply a fiducial scalar $s^{g\kappa}$ and an RMS for its Gaussian
prior.

The density-density cross-correlation $r^{gg}$ may have substantial
impact on cosmological constraints, so we model it with more freedom,
though not full 4-dimensional behavior. We set
\begin{equation}
\label{kggfunc}
s^{gg}_{\alpha \beta i}  =  {\rm max}\left[ 0, 1 - K^{gg}(k,z_i) { |z_\alpha - z_\beta|
    \over 2 \Delta z_{\rm max} } \right].
\end{equation}
This
functional form for $s$ gives the most closely related subsets the
highest covariance.    $\Delta z_{\rm max}$
is the width of the redshift distribution within a set.
The function $K^{gg}(k,z)$ adjusts how quickly the
subsets decorrelate as their photo-z's diverge. 
$K^{gg}(k,z)$ is implemented as the ``$kz$'' functional form described
in Appendix~\ref{functions}, namely a linear
interpolation between a grid of values in the
$(\ln k, \ln a)$ plane.  We hence parameterize the
4-dimensional cross-correlation function by a 2d grid of $K^{gg}$
nodal values.  These points are all given the same fiducial value
$K^{gg}_{\rm fid}$.
The $K^{gg}$ nodal values are given Gaussian priors
to select a range of uncertainty $K^{gg}_{\rm RMS}$.  

The cross-correlation parameters $r^{\kappa\kappa}_{\alpha\beta i}$
are similarly reduced from 4-dimensional behavior by defining 
values $s^{\kappa\kappa}_{\alpha\beta i}$ that are set by the nodal
values of a 2-dimensional function $K^{\kappa\kappa}(k,z)$ in complete
analogy with \eqq{kggfunc}.

\subsection{An apology}
This section on nuisance functions is obscure and lengthy, especially
regarding the cross-correlations of galaxies and intrinsic alignments.
It is, unfortunately, impossible to fully describe the likelihood of
lensing survey data without invoking some model for all of these functions.

Previous work has avoided these messy details and functions by making
many simplifications.
Most have implicitly assumed that all the correlation coefficients are
unity.  Most
have ignored the intrinsic-alignment signal entirely, {\it i.e.}
taking $b^\kappa=0$.  All previous analyses have considered the galaxy
bias to be constant within a set, and if the multiplicative error has
been considered, it has also been constant within a galaxy set.  Only
a few analyses have allowed galaxy bias to vary with redshift
\citep{Zhan06, BJ04}.  The most sophisticated treatment to date
is that of \citet{HJ04}, who take all bias and correlation
coefficients to derive from a halo model of galaxies.  The
redshift-distribution parameters $p_{\alpha i}$ have, in the most
ambitious analyses to date, been reduced to two-parameter (Gaussian)
functions. \citet{MaBernstein} consider sum-of-Gaussian models.  These
assumptions can all be implemented in the present 
formalism if desired, but can also be relaxed to assess their impact
on the cosmological constraints.

\section{Tuning the forecast parameters}
\label{tuneup}
In this section we determine the values of bin widths and
nuisance-function bandwidths that are needed for reliable extraction
of maximum information from lensing surveys.  Unless otherwise noted,
we will derive these parameters for a canonical survey with $f_{\rm
  sky}=0.5$; an effective source density of 60 galaxies per arcmin$^2$
with median redshift of 1.2; $\sigma_\gamma=0.24$; and
Gaussian-distributed fiducial photo-z errors of $\sigma_z=0.04(1+z)$.
Except as noted, we assume $N_{\rm spec}=10^7$ so that photo-z
calibration errors are negligible, and also reduce the shear
calibration RMS uncertainty to $10^{-4}$ to be negligible as well.
Other inputs assume the default values given in
Tables~\ref{params1}--\ref{params5}. 

The information content of a survey will be gauged using the DETF
figure of merit \citep{DETF}: the Fisher matrix will be marginalized
over all nuisance parameters, then the $D_i$ and $\delta F_i$
variables projected onto a model obeying General Relativity with
homogeneous dark energy of equation of state $w=w_0+w_a(1-a)$.  A prior
representing expected Planck results is added (also from the DETF
report), and we marginalize over $\{\omega_m, \omega_b, \omega_k,
\omega_{\rm DE}, n_s, \Delta_\zeta\}$ to yield the Fisher matrix ${\bf
  F}_w$ over $\{w_o, w_a\}$.  The DETF FoM is defined as $|{\bf
  F}_w|^{1/2}$.  
\subsection{Multipole bin size}
Since we have excluded baryon oscillations from our transfer function,
we expect to find little information in the detailed shape of the
lensing or density power spectra.  The broad lensing kernel in
redshift also smoothes away fine structure in the convergence.  So we
expect the information content in the Fisher matrix to be independent of
the multipole bin width $\Delta\log_{10}\ell$ below some modest
value.  Larger values of $\Delta\log_{10}\ell$ reduce the complexity
and execution time of the calculations, so we seek the maximum
$\Delta\log_{10}\ell$ at which nearly all the lensing information is
present.

\begin{figure}
\plottwo{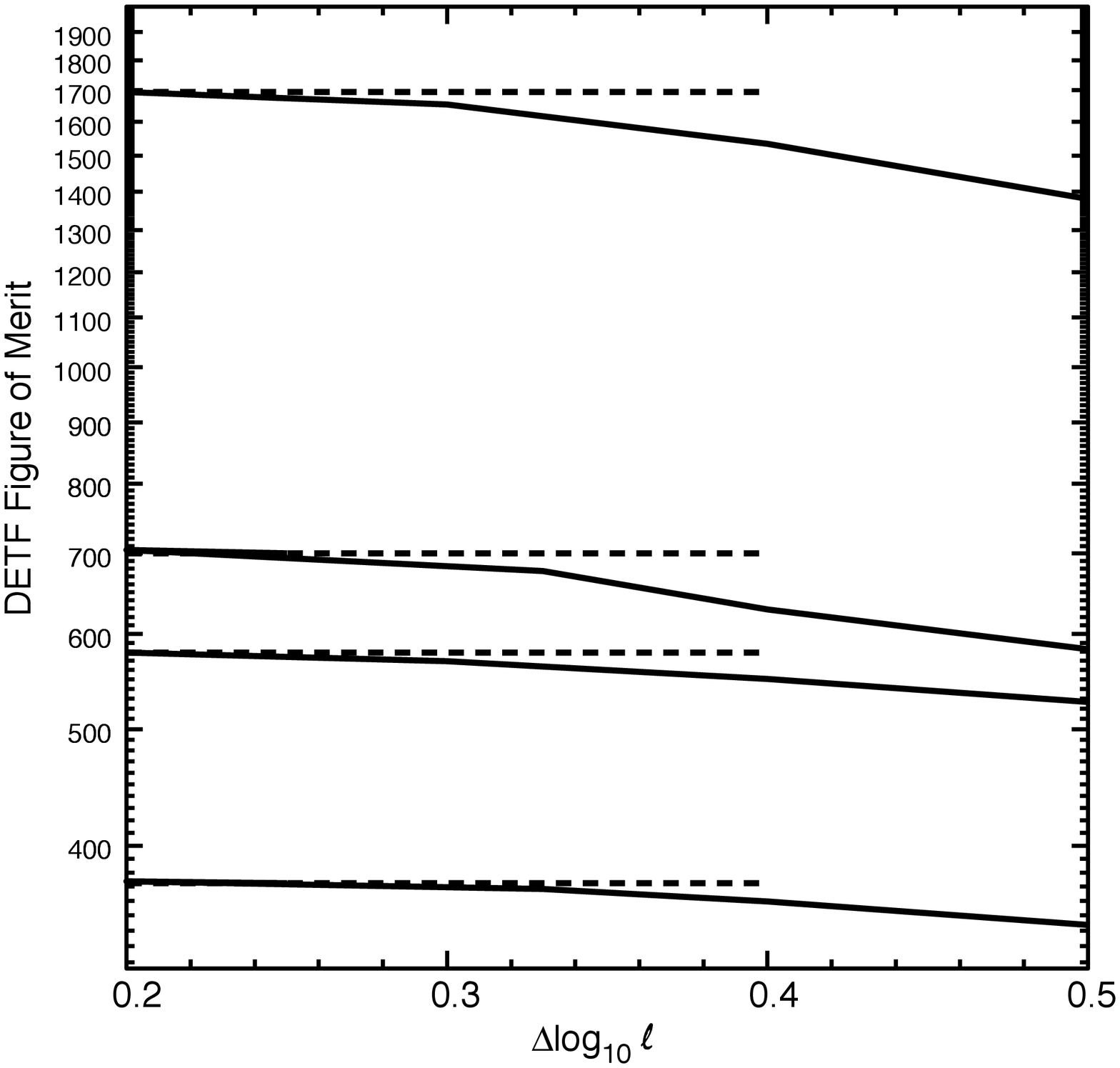}{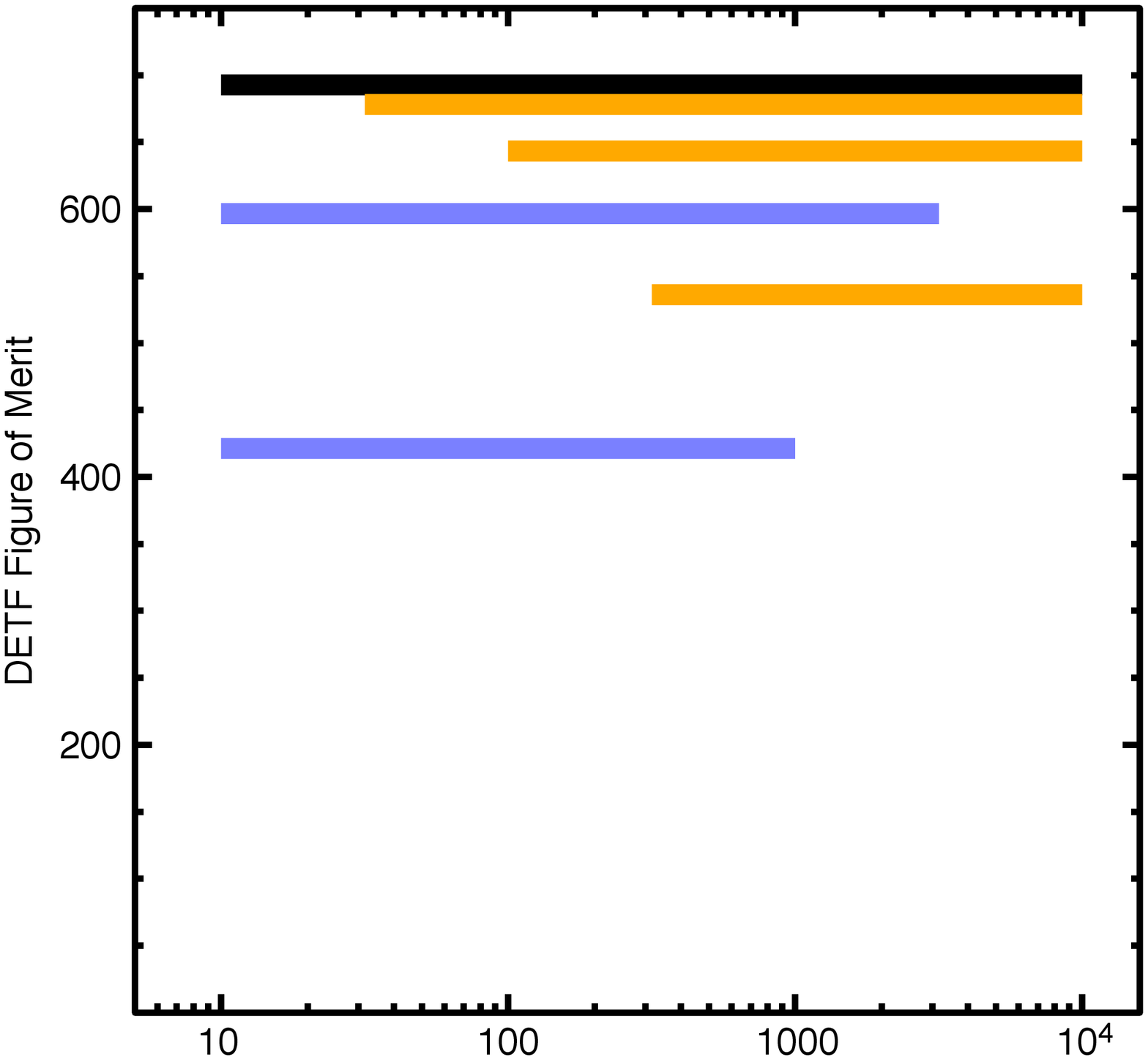}
\caption[]{{\em Left:} DETF Figure of Merit resulting from Fisher analyses of
  several weak lensing surveys, plotted against multipole bin width
  $\Delta\log_{10}\ell$ of the analysis.  Each solid line plots a
  particular survey scenario (see text for details).  The dashed lines
  are horizontal, to 
  help the eye judge the information degradation as we increase
  $\Delta\log_{10}\ell$.  We conclude that $\Delta\log_{10}\ell\le
  0.3$ retains $>97\%$ of the information for surveys of any quality
  level. {\em Right:} Value of DETF FoM for the default survey vs
  range of multipole used.  Choice of upper bound is more critical
  than choice of lower bound on $\ell$.
}
\label{dlogl}
\end{figure}

Figure~\ref{dlogl} plots the DETF FoM of the lensing$+$density survey
(plus spectroscopic redshift survey and Planck prior) vs
$\Delta\log_{10}\ell$ for several candidate surveys. The top line is
for a very optimistic survey: $n_{\rm eff}=100\,{\rm arcmin}^{-2}$,
$N_{\rm spec}=10^7$, $q_{\rm RMS}=10^{-3}$, $f_{\rm RMS}=10^{-4}$,
$b^g_{\rm RMS}=r^g_{\rm RMS}=0.01$, and $b^\kappa_{\rm fid}=10^{-3}$.
By reducing the systematics and the shot noise to (unrealistically)
low levels, we give the lensing survey the chance to extract maximal
information.  We find that the FoM gains only 2\% for
$\Delta\log_{10}\ell<0.3$. 

Other lines in the plot are FoM vs $\Delta\log_{10}\ell$ for weaker
surveys, with $N_{\rm spec}=10^{4.5}$ and/or  $n_{\rm eff}=60\,{\rm arcmin}^{-2}$,
$q_{\rm RMS}=0.1$, $b^g_{\rm RMS}=r^g_{\rm RMS}=0.1$, $b^\kappa_{\rm
  fid}=-0.003$.  In these cases we also find that the DETF FoM
increases by $<2$--3\% for $\Delta\log_{10}\ell<0.3$.

We adopt $\Delta\log_{10}\ell=0.3$ for all future use.

\subsection{Multipole range}
On the right-hand side of Figure~\ref{dlogl} we plot the DETF FoM for
various ranges of $\ell$.  In this study we assume a space based
survey obtaining $n_{\rm eff}=60\,{\rm arcmin}^{-2}$ over $f_{\rm
  sky}=0.5$.  The photo-z and shear calibration systematics are held
negligible with priors but other systematics (intrinsic alignment,
etc.) have default priors.

We see that the choice of $\ell_{\rm max}$ has a strong influence, as
moving from $10^3$ to $10^4$ changes the FoM by $1.7\times$.  We note
this is true even though we have included uncertainty in the
theoretical power spectrum at high $k$ values, showing that there is
still information to be gained when the theory is incomplete.  We
find, in fact, that our default power-spectrum theory uncertainty of
$f_{\rm Zhan}=0.5$ leads to only 6\% degradation of the DETF FoM
relative to
an assumption of {\em zero} uncertainty in the theory, even when
$\ell_{\rm max}=10^4$.  

Unfortunately our assumption of Gaussian statistics will fail by
$\ell=10^4$ \citet{CoorayHu01,LeePen}, rendering the Fisher calculation less
reliable.  We will restrict our analysis to
$\ell<10^{3.5}$, but additional study of the effect of
non-Gaussian statistics is clearly needed.

 The flat-sky and Limber approximations will fail
at low $\ell$, but the choice of $\ell_{\rm min}$ appears less
critical to the $w_0/w_a$ information content, so will retain the
$\ell>10$ bound in our analyses.

\subsection{Scale resolution for nuisance functions}
We require choice of node spacing $\Delta\ln k$ in the nuisance
functions for the power-spectrum theory errors, the calibration errors
$f$ and $q$, and the bias/correlation parameters $b^g_{\rm fine}, r^g,
b^\kappa, r^\kappa, K^{gg},$ and $K^{\kappa\kappa}$.  We set
$\Delta\ln k$ to be equal for all nuisance functions, and find the
value which minimizes the DETF FoM as the ``most damaging'' scale of
variation.  We examine the default case described above, for several values
of the
shear calibration prior $f_{\rm RMS}$ and photo-z calibration size
$N_{\rm spec}$.

\begin{figure}
\plottwo{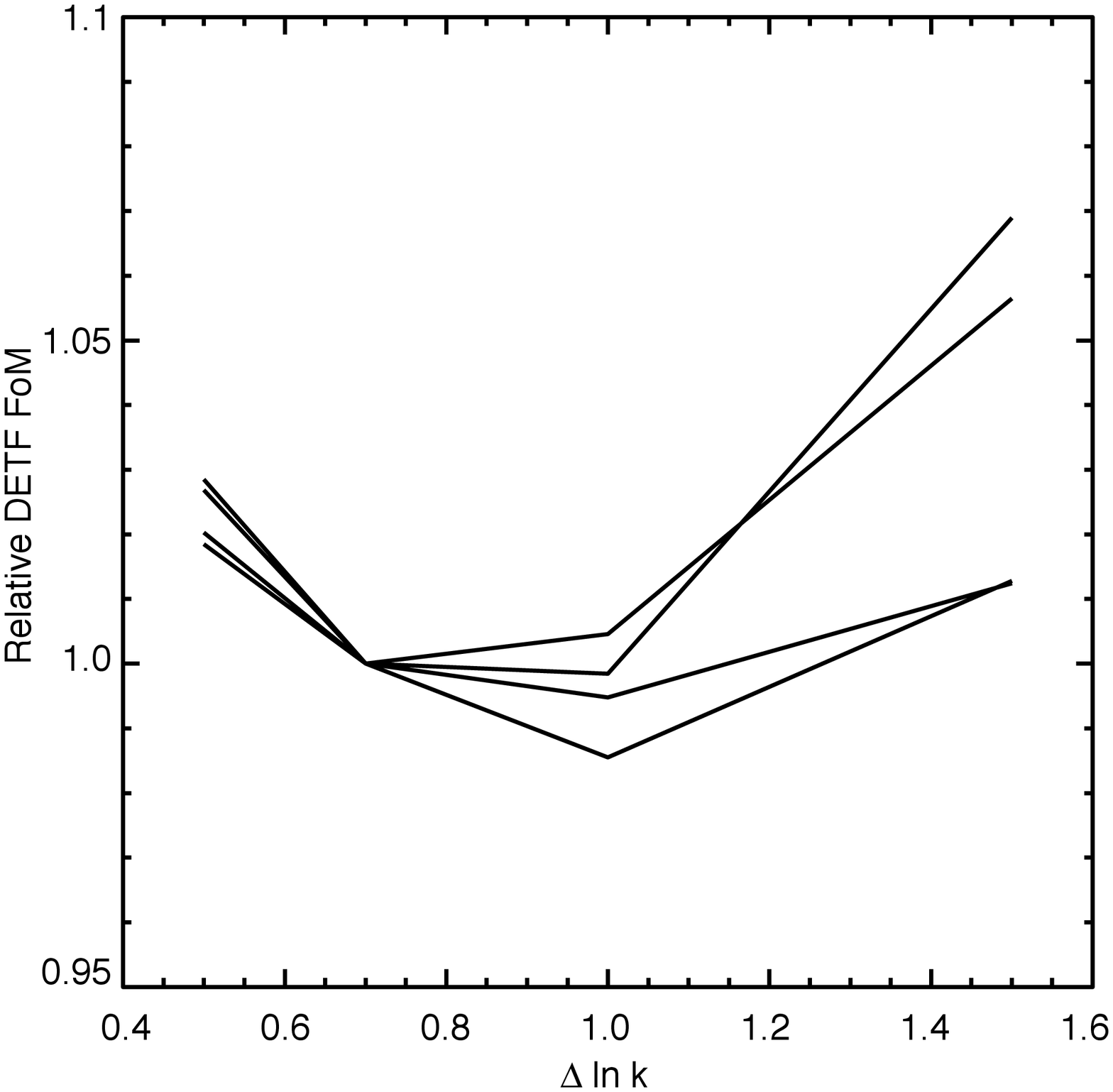}{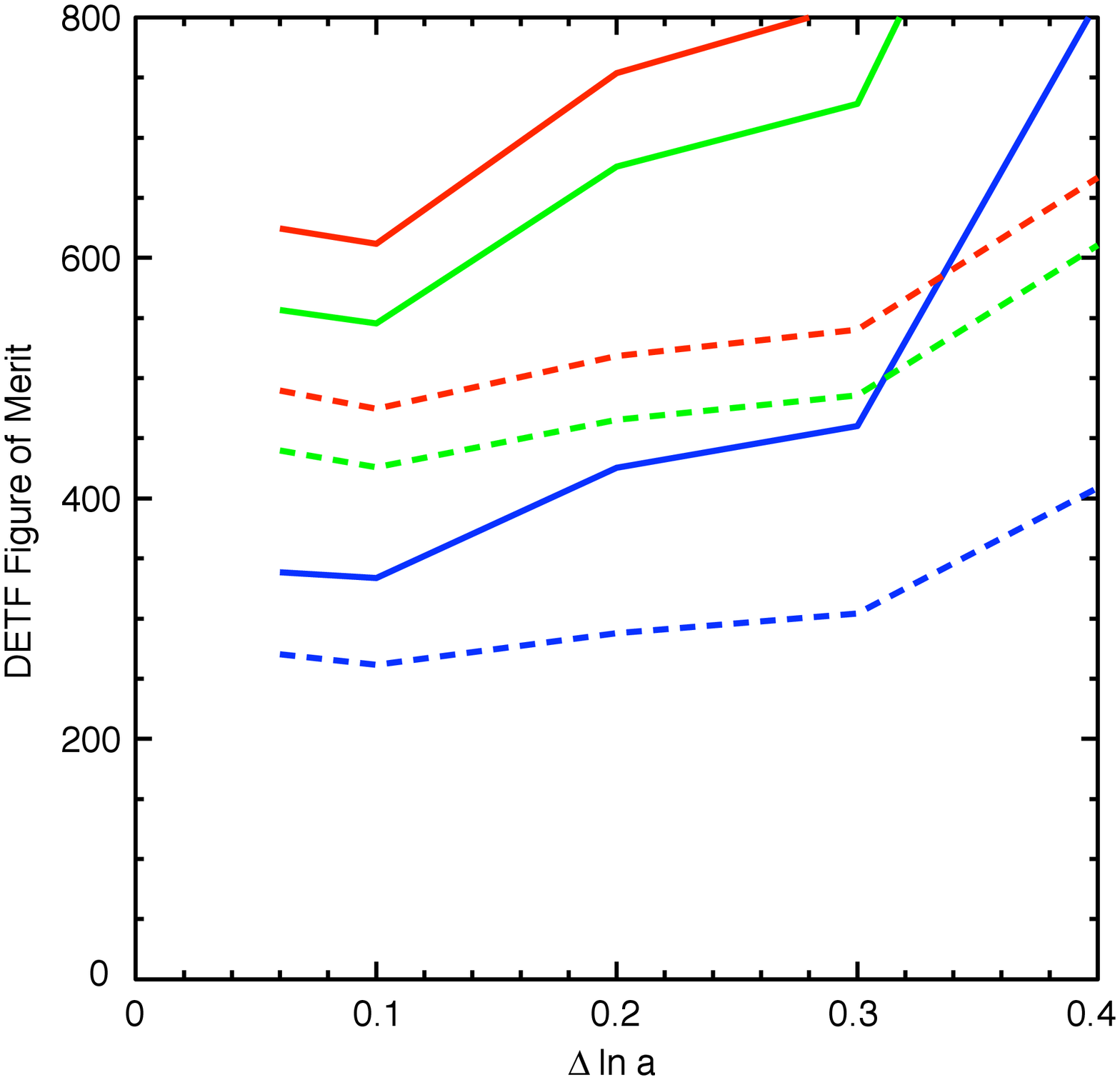}
\caption[]{{\em Left:} DETF FoM vs $\Delta\ln k$, the spacing of nuisance-function
  nodes in the length-scale axis.  We examine surveys with varying
  strengths of priors on shear calibration and photo-z calibration.
  For this plot, each is normalized to the value at $\Delta\ln k=0.7$
  in order to show the (weak) dependence of FoM on $\Delta\ln k$ when
  the prior on RMS nuisance-function fluctuations are held fixed.  We
  adopt $\Delta\ln k=1$ as the most conservative bandwidth for
  nuisance-function variation with scale. {\em Right:} DETF FoM vs
  $\Delta \ln a$, the spacing of nodes in redshift for the bias,
  correlation, and intrinsic-alignment nuisance functions.  The FoM is
  quite sensitive to this choice, and allowing the bias/IA to vary on
  $\Delta\ln a=0.1$ scales is most damaging to cosmological inference.
}
\label{dlnk}
\end{figure}

Figure~\ref{dlnk} shows $\Delta\ln k\approx 0.7$--1 yields minimum
information for fixed RMS priors, but the dependence is very weak.
The FoM varies by only 7\% over the range $0.5<\Delta\ln k<1.5$.  We
henceforth adopt $\Delta\ln k=1$.  Perhaps not surprisingly, this
makes the nuisance functions have $\approx 1$ independent node in each
multipole bin of $\Delta\log_{10}\ell=0.3$.

\subsection{Redshift resolution for nuisance functions}
All of the nuisance functions are dependent on $z$.  We next
investigate the redshift node spacing $\Delta\ln a$ at which the
nuisance functions are most damaging to the DETF FoM.  We find that
the FoM is insensitive to the $\Delta\ln a$ of the power-spectrum
theory errors.  The value of $\Delta\ln a$ for the calibration
functions $f$ and $q$ that minimizes the FoM depends upon the strength
of the prior.  The choice $\Delta\ln a=0.5$ produces a FoM
that is within 2\% of the minimum, however, so we fix this value for
the theory-error and calibration nuisance functions.

The redshift freedom given to the bias and intrinsic-alignment
nuisance functions has a strong impact on the DETF information
content.  Figure~\ref{dlnk} illustrates that, at fixed RMS prior
variation, models with freedom to vary on rather fine scales,
$\Delta\ln a = 0.1$ are most damaging to cosmological
information.

\section{Conclusion}
The core of this paper are the expressions (\ref{gg1})--(\ref{kk1})
for the two-point 
correlation matrix of the lensing and density observable multipoles produced by
a typical lensing survey.  This was derived under a very limited set
of assumptions: a homogeneous and isotropic 4-dimensional metric
Universe with scalar perturbations; plus the weak-lensing limit,
the Limber and Born approximations, and an approximation that lensing
magnification bias and intrinsic density fluctuations are additive.
The last four assumptions could be relaxed at the expense of
computational complexity.  We thus hope that data analyses based on
this framework could be used to constrain a wide variety of potential
explanations for the acceleration phenomenon, including gravity
modifications as well as new fields in the Universe.  In the limit of
Gaussian fluctuation fields, the two-point information is a complete
description of the likelihood and hence can be used to construct
Fisher matrices or analyze data.
As currently configured, the analysis yields the survey's ability to
constrain the distance function $D(z)$ and linear growth function
$g_\phi(z)$, without reference to particular dark-energy models.  It
would be straightforward to implement scale-dependent 
linear-growth functions.

This framework subsumes all of the
information (up to 2-point level) that is likely to be obtained from
lensing observations: density-density, lensing-density, and
lensing-lensing correlations, plus redshift distributions from
unbiased spectroscopic surveys (\S\ref{speclike}).  Furthermore it
allows for the most 
important expected forms of systematic error: photo-z calibration
errors, shear and magnification-bias calibration errors, intrinsic
alignments, and inaccuracies in power-spectrum theory.  Systematics
that are additive to shear (\eg uncorrected PSF ellipticity) or to
density (\eg uncorrected foreground extinction) have not been
included.  We have not done so since the additive errors could, in
principle, exhibit almost any arbitrary signature in the covariance
matrix of the observables.  Hence a completely general model for
additive errors would be degenerate with almost all other signals.
For the additive systematics, it is better to determine the level at
which they would bias the cosmological results than to attempt to fit
a model.  \citet{AmaraRefregier} is a good example of this approach.

Since the analysis framework is independent of models for dark energy,
gravity, power-spectrum evolution, or galaxy bias, we get a
stripped-down look at what parameters are truly constrained by the
data, and what nuisance functions must be modeled in order to extract
the cosmological information.  There is a substantial suite of biases
and correlation functions involved in understanding the full survey
data.  In other work these have been ignored, or have been quantified
by reference to halo occupation models \citep{HJ04,CvdBMLMY}.
Here we introduce generic functions for bias and calibration nuisance
functions that are not based on any
particular physical model.  

We implement one possible model for the evolution of the
lensing-potential power spectrum, based on General Relativity but
allowing for failure of the growth equation.  It is straightforward to
implement other potential deviations from General Relativity.  In the
current implementation, the end result of the Fisher analysis is a
forecast of the ability to constrain the functions $D_A(z)$ and
$g_\phi(z)$.

Since the analysis must be discretized in redshift and angular scale
in order to be feasible, we investigated the bin sizes or bandwidths
of nuisance functions that should be chosen.  We find that $\approx 3$
bins per decade of angular scale suffice to extract all information
(apart from baryon acoustic oscillations), and that nuisance functions
should be specified no finer than this.  Nuisance functions for
power-spectrum theory errors and for shear and magnification-bias
calibration errors can be specified coarsely in redshift space
($\Delta \ln a \approx 1$), but the galaxy biases, correlations, and
intrinsic-alignments must be modeled with potentially finer structure
in redshift ($\Delta \ln a \approx 0.1$) to immunize against potential
astrophysical systematics.

In future papers we will use this framework and its implementation to
investigate the requirements for spectroscopic calibration of
photo-z's in large lensing surveys, and other practical issues.  As a
simple first application of our framework, we have shown here that
power-spectrum theory uncertainty does not significantly degrade the
cosmological power of a nominal lensing survey at $10<\ell<10^4$.
Non-Gaussian statistics are a much more important factor to consider.

{\tt C++} Code to implement Fisher forecasting using this framework has been
written and runs quickly on desktop computers despite the large number
of free parameters in these general models.  Interested parties should
contact the author for access to the code.

\acknowledgments
This work is supported in this work by grant AST-0607667 from the
National Science Foundation, Department of Energy grant
DOE-DE-FG02-95ER40893, and NASA grant 
BEFS-04-0014-0018.  I thank Bhuvnesh Jain, Zhaoming Ma, Chris Hirata,
Ravi Sheth, Hu Zhan,
and the members of the Dark Energy Task Force for helpful
conversations during the long gestation of this work.

\appendix
\section{Parametric functional forms for nuisance variables}
\label{functions}
In modeling an experiment, we often encounter some systematic error
associated with a nuisance variable $f$ about which we have little {\it a
  priori} knowledge.  We would like to fit some parametric form to
this variable, but would like a form that is flexible enough to
describe any ``reasonable'' behavior the function might exhibit.  We
also want to conveniently relate the number and prior probabilities
for the parameters to the kind of variation that $f$ might exhibit.
A parametric description of some nuisance function $f$ defined over a variable
$x\in[-1,1]$ would ideally have the following properties:
\begin{enumerate}
\item $f(x)$ has a variable number $N$ of controlling parameters $\{a_0,
  a_2, \ldots, a_{N-1}\}$ such that any continuous differentiable function
  $F(x)$ can be approximated to any desired accuracy with a
  sufficiently large choice of $N$.
\item We can draw $\{a_j\}$ from independent
  Gaussian distributions of zero mean and widths $\{\sigma_j\}$, with the
  result that ${\rm Var}[f(x)]$ is
  independent of $x$.  In other words, the nuisance value $f$ has a
  uniform and well-determined variance when we apply a simple diagonal
  Gaussian prior to the parameter set $\{a_j\}$.
\end{enumerate}
A Fourier decomposition, $f = \sum (a_j \sin j\pi x + b_j \cos j\pi
x)$, exhibits these qualities, but converges poorly when $f(-1)\ne
f(+1)$.  

\subsection{Linearly interpolated functions}
Another approach is linear interpolation: choosing a spacing $\Delta x
= 2/(N-1)$, we define $a_i$ as the value of $f$ at $x_i=i\Delta x
-1$.  At some other $x_i<x<x_{i+1}$, we define
\begin{equation}
f(x) = wa_i + (1-w) a_{i+1}, \qquad w={x_{i+1} - x \over \Delta x}.
\end{equation}
If we assign an independent Gaussian prior of width $\sigma_a$ to each
$a_i$, then by definition we have ${\rm Var} [f(x)] = \sigma_a^2$ if $x$
coincides with a node.  But the variance of $f$ is not quite
homogeneous: it drops to ${\rm Var} [f(x)] = \sigma_a^2/2$ when $x$ is
halfway between two nodes.  If we want the {\em mean} variance of
$f(x)$ over the interval $x\in[-1,1]$ to equal $\sigma_f^2$, then the
variance of the prior on each node needs to be $\sigma_a^2=3\sigma_f^2/2$.

\subsection{Legendre polynomials}
Polynomial expansions are also commonly used to model nuisance
functions. The simplistic form $f(x)=\sum a_i x^i$ results in
extremely non-uniform variance for $f$ with diagonal prior on
$\{a_i\}$ and is hence inappropriate for our purpose.  A better
choice is to expand in Legendre polynomials $P_n(x)$, which are orthogonal over
$[-1,1]$.  We define
\begin{equation}
f(x) = \sum_{i=0}^{N-1} a_i P_i(\nu x).
\end{equation}
Recall that the Legendre polynomials satisfy $P_0=1, P_1=x,
(n+1)P_{n+1}=(2n+1)xP_n-nP_{n-1}$.

We wish to choose priors $\{\sigma_i\}$ on $\{a_i\}$ that cause
the variance of $f(x)$ be as uniform as possible for
$x\in[-1,1]$.  We have not found a way to attain perfect uniformity in $x$
with polynomial interpolation, however the following scheme gets
usefully close.  We discover numerically that
\begin{equation}
\lim_{N\rightarrow\infty} {1 \over N} \sum_{i=0}^{N-1} P_n^2(\nu x) (2i+1)
= {2 \over \pi} \left(1-\nu^2x^2\right)^{-1/2}.
\end{equation}
This implies that, if we set $\sigma_i = \sqrt{(2i+1)\pi/2N}$, then in
the limit of large $N$ we will obtain ${\rm Var} [f(x)] =
\left(1-\nu^2x^2\right)^{-1/2}.$  For $\nu=1$ this would diverge at
the ends of our nuisance function's interval.  If, however, we choose
$\nu=0.9$, the RMS is only $\approx1.5\times$ larger at the
endpoints than at $x=0$.  Over the $[-1,1]$ interval, the mean
variance is $(\sin^{-1} \nu)/\nu$.  Hence if we wish to have a
function with $\langle {\rm Var} f(x) \rangle_x = \sigma^2_f$, we set
the priors on the Legendre coefficients to be
\begin{equation}
\sigma_i = \sigma_f \sqrt{(2i+1)\pi \nu \over 2N \sin^{-1}\nu}.
\end{equation}

\subsection{Standard multidimensional functions}
The Fourier, linear-interpolation, and Legendre-polynomial functional
forms can each be extended to $>1$ dimensions in straightforward
fashion.  In the lens modeling, we need functions of these
dimensions:
\begin{enumerate}
\item Comoving wavenumber or physical scale: $x_1=\ln k$;
\item Redshift $z$: more precisely we will use the variable $x_2=\ln
  (1+z)$;
\item Photometric redshift error $\Delta z$; more precisely, our code
  uses the variable $x_3 = \Delta \ln (1+z) = \ln[(1+z_\alpha)/(1+z_i)]$ for
  subset $\alpha i$.
\end{enumerate}

\subsubsection{$kz$ form}
For functions over the $(x_1,x_2)$ space we use a simple
two-dimensional version of interpolation between values on a
rectangular grid.  The power-spectrum theory error $\delta\ln P$ uses
this form, as do the $K^{gg}(k,z)$ and $K^{\kappa\kappa}(k,z)$
nuisance functions.  The only complication of note is that the nodal
point $a_{ij}$ should have a prior with variance
$\sigma^2_a=(3/2)^2\sigma^2_f$ if the output function is to have
variance $\sigma^2_f$.  The factor of $(3/2)^2$ is needed to
counteract the reduced variance when interpolating between grid points.

The $kz$ nuisance function is specified by:
\begin{itemize}
\item the spacing $\Delta x_1 = \Delta \ln k$ of the nodes for
  linearly interpolation in $x_1$;
\item the spacing $\Delta x_2 = \Delta \ln (1+z)$ of the nodes for
  linearly interpolation in $x_2$;
\item the RMS variation $\sigma_f$ of the function 
  allowed under the prior, which can depend on $x_1$ and $x_2$.
\item the fiducial dependence of $f$ on $x_1$ and $x_2$.
\end{itemize}

\subsubsection{$z\Delta z$ form}
For nuisance variables over the $(x_2,x_3)$ space ($z$ and $\Delta
z$), we adopt the following strategy:
we choose to have variation over $x_3$ be described by
polynomials since we usually define the range of non-catastrophic
photo-z errors to be 
bounded to some range $|x_3| \le \Delta_{\rm max}$.  We define
the ``$z\Delta z$'' functional form as follows:
\begin{equation}
f(x_2,x_3) = \sum_{i=0}^{N-1} a_i(x_2) P_i(\nu x_3 / \Delta_{\rm max}).
\end{equation}
The Legendre coefficients are in turn defined to be linearly
interpolated between a series of values $a_{ij}$ at redshift nodes
$\{\ln (1+z_j)\}$.  The $a_{ij}$ become parameters of the model.  

The fiducial and prior values for the $i=0$ terms (constant in $x_3$)
are treated differently than the $i>0$ terms.  We might, for
example, expect some nuisance functions to vary strongly with $x_2$
(nominal redshift) but only slightly with $x_3$ (photo-z error) at
fixed $x_2$.  

We specify the total RMS fluctuation in $f$ allowed
by the prior to be $\sigma_f$. But we also specify the fraction {\tt
  VarFracDZ} of the variance that is due to dependence on $x_3$. If
we define
\begin{equation}
\bar f(x_2) \equiv \int_{-\Delta_{\rm max}}^{+\Delta_{\rm max}} dx_3\,f(x_2,x_3) /
2\Delta_{\rm max},
\end{equation}
then we aim to achieve
\begin{eqnarray}
{\rm Var} [\bar f(x_2)] & = & \sigma^2_f \left(1-{\tt VarFracDZ}\right) \\
{\rm Var} \left[f(x_2,x_3)-\bar f(x_2)\right] & = & \sigma^2_f \left({\tt
    VarFracDZ}\right).
\end{eqnarray}

This is achieved approximately by setting the priors on the constant terms as
\begin{equation}
\label{ap1}
\sigma^2_{0j} = \frac{3}{2} \sigma^2_f \left(1-{\tt VarFracDZ}\right) 
\end{equation}
and the $x_3$-dependent terms as
\begin{equation}
\label{ap2}
\sigma^2_{ij} = \frac{3}{2} \sigma^2_f \left({\tt VarFracDZ}\right) 
\sqrt{(2i+1)\pi \nu \over 2N \sin^{-1}\nu}.
\end{equation}

The RMS prior variation $\sigma_f$ can be made a function of $x_2$
without loss of generality.
We typically take fiducial values of $a_{ij}=0$ for $i>0$ in our
nuisance functions, \ie no fiducial dependence upon $\Delta z$.

To summarize, the $z\Delta z$ nuisance function is specified by:
\begin{itemize}
\item the order $N$ of the polynomial in $x_3$;
\item the maximum range $\Delta_{\rm max}$ of applicability in the
  $x_3$ axis;
\item the spacing $\Delta x_2 = \Delta \ln (1+z)$ of the nodes for
  linearly interpolation in $x_2$;
\item the RMS variation $\sigma_f$ of the function allowed under the
  prior, which can depend on $x_2$;
\item the fraction {\tt VarFracDZ} of this variance that is due to
  $x_3$ ($\Delta z$) dependence;
\item the fiducial dependence of $f$ on $x_2$.
\end{itemize}

\subsubsection{$kz\Delta z$ form}
The bias and correlation coefficients can, most generally, depend on
scale ($x_1$) as well as subset $(x_2,x_3)$, so we generalize to the
``$kz\Delta z$'' functional form:
\begin{equation}
f(x_1,x_2,x_3) = \sum a_i(x_1,x_2) P_i(x_3).
\end{equation}
The coefficients
$a_i$ are linearly interpolated
between nodes in the two-dimensional space $(x_1,x_2)$.  Thus the free
parameters of this model become the Legendre coefficients $a_{ijk}$.
As for the $z\Delta z$ function, we specify the prior by the overall
mean RMS variation $\sigma_f$ (which can be a function of $x_2$ and
$x_3$), plus the {\tt VarFracDZ} specifying how much of the variance
is manifested as dependence on $\Delta z$.  The formulae for the
priors $\sigma_{ijk}$ on the nodal coefficients are derived exactly as
in Equations~(\ref{ap1}) and (\ref{ap2}), except that we now need
factors of $(3/2)^2$ to account for the reduced variance when
interpolating in two dimensions.

The $kz\Delta z$ function thus requires all of the specifications as
the $z\Delta z$ function, plus a spacing $\Delta\ln k$ for nodes in
the $x_1$ axis.

\end{document}